\documentclass[useAMS,usenatbib]{mn2e}

\usepackage{graphicx} 
\usepackage{amsmath}

\title[Multidimensional modelling of Cepheid-like variables. I:~Extensions of the ANTARES 
code.]{Multidimensional realistic modelling of Cepheid-like variables. I:~Extensions of the ANTARES code.}
\author[E. Mundprecht, H.J. Muthsam and F. Kupka]{Eva Mundprecht$^{1}\thanks{E-mail:
eva.mundprecht@univie.ac.at}$,  Herbert J. Muthsam$^{1}\thanks{E-mail:
herbert.muthsam@univie.ac.at}$ and Friedrich Kupka$^{1}$\\
$^{1}$University of Vienna, Faculty of Mathematics, Nordbergstra{\ss}e 15, A-1090 Wien, Austria}

\begin{document}

\date{Accepted ..... Received ......; in original form...}

\pagerange{\pageref{firstpage}--\pageref{lastpage}} \pubyear{xxxx}

\maketitle

\label{firstpage}

\begin{abstract}
We have extended the ANTARES code to simulate the coupling of pulsation with convection
in Cepheid-like variables in an increasingly realistic way, in particular in multidimensions, 2D 
at this stage. Present days models of radially pulsating stars assume radial symmetry and 
have the pulsation-convection interaction included via model equations containing ad hoc 
closures and  moreover parameters whose values are barely known. We intend to construct 
ever more realistic multidimensional models of Cepheids. In the present paper, the first of 
a series, we describe the basic numerical approach and how it is motivated by physical properties
of these objects which are sometimes more, sometimes less obvious. -- For the 
construction of  appropriate models a polar grid co-moving with the mean radial velocity has been introduced
to optimize radial resolution throughout the different pulsation phases. The grid is radially stretched 
to account for the change of spatial scales due to vertical stratification and a new grid refinement 
scheme is introduced to resolve the upper, hydrogen ionisation zone where the gradient of temperature
is steepest. We demonstrate that the simulations are not conservative when the original weighted
essentially non-oscillatory method implemented in ANTARES is used and derive a new scheme
which allows a conservative time evolution. The numerical approximation of diffusion follows the
same principles. Moreover, the radiative transfer solver has been modified to improve the efficiency 
of calculations on parallel computers. We show that with these improvements the ANTARES code 
can be used for realistic simulations of the convection-pulsation interaction in Cepheids. We 
discuss the properties of several numerical models of this kind which include the upper 42\% of 
a Cepheid along its radial coordinate and assume different opening angles. The models are 
suitable for an in-depth study of convection and pulsation in these objects.
\end{abstract}


\section{Introduction}

The series of papers introduced by this article deals with the construction and analysis of time-dependent, 
nonlinear numerical models for classical, radially pulsating variable stars (Cepheids, RR Lyr,...)
in multiple dimensions. The models are set up to investigate the pulsation-convection interaction and, eventually, 
nonradial pulsations in such stars. In the present paper we describe the extension of the ANTARES code 
\citep{muth_na10} necessary for such an investigation. We additionally discuss effects of resolution and
other issues. A subsequent paper \citep{muth_mnras12}, paper~II, will deal with an investigation of the physical results. 

The following considerations have motivated this work. One of the early successes of numerical modelling of the 
dynamics of astrophysical objects were simulations of Cepheid (or RR Lyr, W Vir,...) pulsations. It was one of the 
few time-dependent nonlinear problems in astrophysics where the approximation of perfect spherical symmetry was 
sufficiently good while at the same time observations at the dynamical timescale were readily available. Even decades
earlier systematic trends of the shapes of the lightcurve with period, the famous Hertzsprung progression
in classical Cepheids, had been established (\citealt{hertz26}). Interest in the subject continued to be kindled by the
importance of these variable stars for the understanding of stellar evolution, in particular when discrepancies 
between masses derived from stellar evolution considerations and from pulsation properties were found
(\citealt{cox80} and \citealt{keller08}). The decisive role of Cepheids in gauging the cosmic distance scale is
well known (see the extensive discussion in \citealt{degrijs11}, e.g.) and the precise calibration of relations between
mass, luminosity, and metallicity for different types of classical variables is of general astrophysical interest.
 
Linear stability considerations suggested the excitations of the pulsations through the ionisation of hydrogen and, 
in particular, helium (\citealt{zhev53}, \citealt{cox60}, \citealt{bak62}).
The first nonlinear, time-dependent calculations soon followed.
For pioneering nonlinear work see \cite{christy62}, \cite{aleshin64},  \cite{christy64}, and \cite{cox66}. 

Spherical symmetry allowed the problem to be treated computationally in one spatial dimension and 
made it accessible to the computing resources of that period. The subsequent decades, essentially till now, have 
witnessed substantial refinements of computational methods and input physics. Let us mention the inclusion of radiative 
transfer for the outer zones instead of the diffusion approximation, better numerics including adaptive grids, more realistic 
opacities and the like. There now exists a considerable number of codes of that kind with various properties such as described 
in  \cite{fok90}, \cite{dorfi91}, \cite{bono_st94}, \cite{buch_koll97} and \cite{smolec_mo08}.

It soon became apparent that in order to properly represent important observational facts such as the red edge of the 
Cepheid instability strip convection had to be included and that attempts to accomplish this by incorporating standard
mixing length theory into a pulsation code failed \citep{tuggle_iben73}. Consequently, practically all groups working on the
nonlinear hydrodynamics of pulsating stars included convection models which from the outset were meant to apply to
a changing environment, essentially extensions of the  mixing length approach or models for the time evolution
of the turbulent kinetic energy (TKE). 

While this improved matters in many respects, it certainly did not solve the problem of constructing an 
accurate, predictive model of the convection-pulsation interaction in these stars. Since such models
cannot be derived from basic principles alone, a variety of recipes for modelling the convective flux is in use 
(and is discussed in the next paper of this series). The same holds for many other physical quantities appearing 
in these models. Many closure constants, up to eight, ``the $\alpha\mbox{'s}$'', are needed to fully specify
the model equations currently used for the one-dimensional calculations (see the impressive compilation 
in \citealt{buchko00}). Their values are unknown and must be determined somehow. 
Common ways to do so include adjusting the parameters to match some observable quantities or simply 
guessing them. Consequently, even if observed properties are reproduced, the extent to which this success 
can be justifiably ascribed to the model and does not constitute a mere result of the fitting procedure remains 
doubtful. It most likely lacks the deeper physical correctness intended to achieve with unknown consequences 
for further interpretations based on such a model.

The state of knowledge has recently been discussed in the review by \cite{buch_rev_09}. There, he addresses
the use of time dependent mixing length theory as the ``dominant deficiency'' in Cepheid modelling (see also
\cite{buch_rev97}). Again in a review, \cite{marc_rev09} emphasized the proper inclusion of convection for 
modelling RR Lyr stars.

Recently, activities have commenced to address this issue. \cite{gast_L11} thoroughly analyse physically
simplified 2D models to elucidate basic  mechanisms of the pulsation-convection interaction (see also
literature cited therein). \cite{ger11} have reassumed very early work by \cite{deup80}. Their calculations are
in 3D, albeit with quite low resolution at present. Various physical simplifications have been made in this work
(no realistic microphysics, adiabatic stratification), but more realistic microphysics, opacities, and an energy 
equation based on the diffusion approximation are intended to be included in future investigations.

This is the context in which we present the extension of the ANTARES code for the type of research at hand.
We extend the main features of the code (radiation-hydrodynamics with radiative transfer equations in the
outer regions, realistic microphysics and opacities, high-resolution numerics, optional grid-refinement) as 
required for work on nonlinear stellar pulsation. We furthermore add a moving grid (moving at the upper 
boundary with the mean vertical surface  velocity). While the main part of ANTARES is designed for the 
1D, 2D, and 3D case, for our present work only the 1D and 2D case have been implemented.
This is due to the fact that the extremely steep gradients near the hydrogen ionisation zone make even
an adequately resolved 2D calculation a computationally very expensive enterprise. Undoubtedly, however,
in the future we will also include the option of performing numerical simulations of this kind in 3D. 

At first sight it may look astonishing that we are going to undertake 2D calculations for the time being.
After all, meaningful 3D models of solar granulation exist since decades \citep{nord82}. For Cepheids, 
however, the circumstances are different. We find a few remarks on the 2D \textit{model atmospheres} of 
Cepheids in \cite{frey12}. The difficulties are basically the same as those encountered in the simulation 
of A-type stars and extensively discussed for them by \cite{kup09} and, for Cepheids, in 
Sect.~\ref{sec:demands} of the present article where also a few remarks on computational demands 
are provided. -- The 3D nature of the simulations presented in \cite{ger11} does not contradict those 
remarks, since their work considers an adiabatic test case whose idealised microphysics lacks the 
properties which ultimately lead to the aforementioned problems.

With the goal of modelling  such stars in multidimensions with an ever increasing degree of realism we have 
extended the ANTARES code and continue to do so, maintaining its basic design principles and ingredients 
as described in \citet{muth_na10}, namely
\begin{itemize}
\item{high resolution numerics of the ENO variety}
\item{realistic equation of state and microphysics}
\item{radiative transfer in the diffusion approximation or 
using the static radiative transfer equations}
\item{parallelization based on MPI and OpenMP.}
\end{itemize} 

For our present specific purpose extensions are necessary which are either obvious from the outset 
or which experience has taught to be indispensable, namely
\begin{itemize}
\item{the use of polar coordinates}
\item{a moving grid to follow the pulsations of the star}
\item{a radially stretched grid for coping with the vast differences  in  
length scales between the stellar surface and the interior}
\item{adaption of the treatment of hydrodynamics and radiative transfer to
these changed paradigms}
\item{specific forms of grid refinement to properly cope with 
the fast spatial variation of quantities in very localized regions }
\item{considerations of how to formulate the numerical scheme given the 
requirements of the objects envisaged for research.}
\end{itemize}

The aim of the paper is, therefore, 
\begin{itemize}
\item to discuss just those numerical considerations and 
how they are brought about by the physics of pulsating stars
\item and to provide an 
overview of what is accessible to multidimensional numerical modelling 
of Cepheids and related variables presently or in the foreseeable future.
\end{itemize}

For that purpose we have organised the remaining sections as follows. 
We first present the equations 
of radiative hydrodynamics on a stretched, co-moving and polar grid and some numerical methods 
adapted to this setting. In Sect.~\ref{sec:demands} we provide the parameters of our model and 
investigate the restrictions resulting from the steep gradient in the H-ionisation zone, while in 
Sect.~\ref{sec:grid} we analyse the effects of resolution. Finally, we discuss the usefulness of
the calculated simulations, their strength and weaknesses, and possible improvements
in Sect.~\ref{sec:discussion} before we provide our conclusions in Sect.~\ref{sec:conclusions}. 

%
 %
\begin{table*}
 \centering
 \begin{minipage}{\textwidth}
\begin{tabular}{|l|l|l|l|l|l|l|l|l|}
\hline
model&aperture& \multicolumn{2}{c}{grid points}   &stretching&subgrid   &grid           &radial&refined radial\tabularnewline
nr.     &angle       &radial         &polar&factor      &modelling&refinement&cell size&cell size\tabularnewline
\hline
1      & 10$^{\mathrm{o}}$ & 510  & 800  & 1.011     & no        & no      & 0.47 \dots 124 Mm  & \tabularnewline
2      & 3$^{\mathrm{o}}$   & 510  & 300  & 1.011     & yes      & no      & 0.47 \dots 124 Mm  & \tabularnewline
3      & 3$^{\mathrm{o}}$   & 510  & 300  & 1.011     & yes      & yes    & 0.47 \dots 124 Mm  & 0.32 \dots 0.80 Mm\tabularnewline
4      & 1$^{\mathrm{o}}$   & 800  & 300  & 1.007     & yes      & no      & 0.29 \dots 79 Mm   &\tabularnewline
\hline
\end{tabular}
\end{minipage}
\caption{\quad{}Grid parameters for the different numerical Cepheid models discussed in this paper.}
\label{tab:models}
\end{table*}

\section{Numerical methods} \label{sec:model}

The simulations of radially pulsating stars are
 performed on a stretched, polar, and moving grid. We first introduce the basic conservation
laws governing the time evolution of our numerical models in Sect.~\ref{subsec:equations} and 
define subsequently the computational grid 
(Sect.~\ref{gridstructure}). In Sect.~\ref{subsec:RTE} the necessary alterations for the radiative transfer 
equation (RTE) are explained in detail. They are not only required for realistic simulations of layers near 
the stellar surface, but also alleviate the restrictions on time 
stepping (see Sect.~\ref{subsec:timestep}). In Sect.~\ref{subsec:WENO} a new set of ENO-coefficients 
is presented that ensures conservation of the hydrodynamic quantities on a stretched polar grid. Similar
adaptations for discretising diffusive fluxes are discussed in Sect.~\ref{subsec:diffusion} while 
Sect.~\ref{subgrid} explains our optional subgrid scale modelling. Finally, Sect.~\ref{initbound} explains
the setup of initial and boundary conditions of the simulations.

\subsection{Conservation equations}
\label{subsec:equations}

Including the grid velocity $\overrightarrow{u_{g}}$ in the conservation equations for mass, momentum, and total (kinetic plus
thermal) energy we obtain as our set of dynamical equations

\begin{equation}
    \frac{\partial\rho}{\partial t}=-\nabla\cdot\left[\vec{I}-\rho\overrightarrow{u_{g}}\right],
\label{eq:denscons}
\end{equation}

\begin{equation}
    \frac{\partial\vec{I}}{\partial t}=-\nabla\cdot\left[\frac{\vec{I}\vec{I}}{\rho}-\vec{I}\overrightarrow{u_{g}}-\overline{\sigma}+p\cdot \mbox{Id}\right]+\frac{source}{r}+\rho\vec{g},
\label{eq:momcons}
\end{equation}

\begin{equation}
   \frac{\partial e}{\partial t}=-\nabla\cdot\left[\frac{\vec{I}}{\rho}\left(e+p\right)-e\overrightarrow{u_{g}}-\frac{\vec{I}} {\rho}\cdot\overline{\sigma}\right]+\vec{g}\cdot\vec{I}+Q_{\rm rad},
\label{eq:enercons}
\end{equation}
where the term containing
\begin{equation}
source=\left(\begin{array}{r}
2p+\frac{I_{\varphi}^{2}+I_{\theta}^{2}}{\rho}\\
\frac{-I_{r}I_{\varphi}+I_{\varphi}I_{\theta}\cot\theta}{\rho}
\\
-p\cot\theta-\frac{I_{r}I_{\theta}+I_{\varphi}^{2}\cot\theta}{\rho}
\end{array}\right)
\end{equation}
originates from introducing spherical coordinates. $\overline{\sigma}$ is the viscous stress tensor, $\rho$ denotes 
density, $\vec{I}=\left(I_{r},I_{\varphi},I_{\theta}\right)$ is the momentum density, $e$ the total energy density, 
$p$ the pressure, and $\vec{g}$ the gravitational acceleration. The grid velocity at the boundaries is set to the 
horizontal fluid velocity average at the top, $u_{g}|_{\mathrm{top}}=\overline{u_{r}}|_{\mathrm{top}}$, and 
zero at the bottom. For the intermediate points 
$u_{g}(i)=\left(r_i-r_{\mathrm{bot}}\right)/\left(r_{\mathrm{top}}-r_{\mathrm{bot}}\right)$.
The radiative heating rate $Q_{\mathrm{rad}}$ is computed by solving the radiative transfer equation (RTE) 
in optically thin regions, and by the diffusion approximation $-\nabla\cdot\left(\chi\nabla T\right)$
elsewhere. $\chi$ is the radiative conductivity. 
The system is closed by an equation of state, which describes the relation between the 
thermodynamical quantities and depends on the physical properties of the fluid. Since the simple 
thermodynamical relations for an ideal gas are not applicable to a partially ionised gas, realistic microphysics is 
included by the LLNL equation of state (\citealt{opalEos}, \citealt{rogers_nay_02}) and OPAL opacities 
(\citealt{opalOpac}). For the RTE the \cite{aleFer} low-temperature Rosseland opacities for grey
radiative transfer are used to extend the accessible temperature range.

\subsection{The grid structure}
\label{gridstructure}

The computational domain is equipped with a 3D spherical grid, which is used for 2D simulations and for 1D simulations,
such that one cell covers one entire shell at a given radius. In radial direction $N_{r}$ grid points
plus 4 ghost cells at each boundary are used. The values of  $N_{r}$ etc. for the different models can be found in
Table~\ref{tab:models}. The $r$-range covers $r\in\left[r_{\mathrm{bot}},r_{\mathrm{top}}\right]$, where
$r_{\mathrm{bot}}$ is fixed and $r_{\mathrm{top}}$ varies with time. The grid is stretched in radial direction
by a factor $q$. The mesh sizes are $\Delta r_{i+1}$=$q\Delta r_{i}$ varying from top to bottom. Thus, the numerical 
grid becomes \begin{equation}
r_{i}=r_{\mathrm{bot}}+\frac{q^{i}-1}{q^{N_{x}}-1}\left(r_{\mathrm{top}}-r_{\mathrm{bot}}\right).
\label{eq:gridStrucRad}
\end{equation}

Cell boundary values are defined as $r_{i+\frac{1}{2}}=r_{i}-\frac{\Delta r_{i}}{1+\sqrt{q}}$ and
$r_{i-\frac{1}{2}}=r_{i}+\frac{\sqrt{q}\Delta r_{i-1}}{1+\sqrt{q}}$. In azimuthal direction the angle $\varphi$ covers
$\varphi_{j\pm\frac{1}{2}} \in
\left[ -\frac{\varphi_{tot}}{2},
        +\frac{\varphi_{tot}}{2} \right]$,
and the mesh size is $\Delta\varphi=\frac{\varphi_{\mathrm{tot}}}{N_{\varphi}}$. The
numerical grid is given by
\begin{equation}
  \varphi_{j}= -\frac{\varphi_{\mathrm{tot}}}{2} + (j-\frac{1}{2})\Delta\varphi.
\label{eq:gridStrucPhi}
\end{equation}
Cell boundary values are defined as $\varphi_{j\pm\frac{1}{2}}=\varphi_{j}\pm\frac{\Delta\varphi}{2}$. The physical
distance between two adjacent points in azimuthal direction is computed as
$\Delta y_{ik}=r_{i}\Delta\varphi\sin{\theta_{k}}$. The polar (colatitudinal) angle covers the range
$\theta_{k\pm\frac{1}{2}} \in
\left[ \frac{\pi-\theta_{tot}}{2},
        \frac{\pi+\theta_{tot}}{2} \right]$,
the mesh size is $\Delta\theta=\frac{\theta_{\mathrm{tot}}}{N_{\theta}}.$ The
numerical grid is given by
\begin{equation}
 \theta_{k}=k\Delta\theta + \frac{\pi-\theta_{\mathrm{tot}} - \Delta\theta}{2}.
\label{eq:gridStrucTheta}
\end{equation}
Cell boundary values are defined as $\theta_{k\pm\frac{1}{2}}=\theta_{k}\pm\frac{\Delta\theta}{2}$. The physical distance
between two adjacent points in colatitudinal direction is computed as $\Delta z_{i}=r_{i}\Delta\theta$.

For 2D simulations there is just one cell of size $\theta_{\mathrm{tot}}=\Delta\theta=\pi$ in colatitudinal direction and the 
grid reduces to  $\theta_{\frac{1}{2}}=0$,  $\theta_{1}=\frac{\pi}{2}$, and $\theta_{\frac{3}{2}}=\pi$ and for 1D simulations
in addition the azimuthal grid becomes $\varphi_{\frac{1}{2}}=-\pi$,  $\varphi_{1}=0$, and $\varphi_{\frac{3}{2}}=\pi$ with 
$\varphi_{\mathrm{tot}}=\Delta\varphi=2\pi$, so that one cell represents one shell at a given radius. 
In 2D simulations we hence consider sectors located in the equatorial plane of the sphere as our simulation boxes
while in 1D the simulation boxes are radial columns.
In ANTARES the $x$-direction of the Cartesian grid points inward, thus for compatibility with various routines 
$x_{i}=r_{\mathrm{top}}-r_{i}$ is used and for visualization purpose there is a Cartesian output grid. There, the 
$x$-coordinates of the radius at $\varphi=0$ and $\theta=\frac{\pi}{2}$ appear negative which has to be 
considered when evaluating and visualizing the output data.

\subsection{The radiative transfer equation (RTE)}
\label{subsec:RTE}

To perform a realistic simulation of layers near the stellar surface the nontrivial exchange of energy
between gas and radiation is included via the RTE, while the diffusion approximation is used for the 
deeper, optically thick layers to reduce computing time. The RTE in 1D 
\begin{equation}
\mu \frac{\partial I}{\partial \tau }=I-S
\end{equation}
is solved along single rays via the short characteristics method of  \cite{kun_auer}.  Here, $I$  denotes the 
intensity, $S$ the source function (taken to be the Planck function), and $\tau$ the vertical optical depth scale. 
Finally, $\mu=\mathrm{cos}\, \theta$, where $\theta$ is the polar angle. In 2D either 12 or 24 ray directions 
are chosen according to the angular quadrature formulae of types A4 or A6 of \cite{carlson}, and the directions 
in each quadrant are arranged in a triangular pattern (Tables \ref{tab:A4} and \ref{tab:A6}). This is a reduction
of a factor of two compared to the 3D case, a result which follows from the additional symmetry.

\begin{table}
\renewcommand{\arraystretch}{1.5}
\begin{tabular}{|c|c|c|c|}
\hline
i & x$_{i}$ & y$_{i}$ & z$_{i}$ \tabularnewline
\hline
1 & $\frac{1}{3}$ & $\frac{1}{3}$ & $\sqrt{\frac{7}{9}}$ \tabularnewline
2 & $\frac{1}{3}$ & $\sqrt{\frac{7}{9}}$ & $\frac{1}{3}$ \tabularnewline
3 & $\sqrt{\frac{7}{9}}$ & $\frac{1}{3}$ & $\frac{1}{3}$ \tabularnewline
\hline
\end{tabular}
\caption{\quad{}Quadrature formula Carlson A4 for integration over a sphere: ray directions in the first octant.}
\label{tab:A4}
\end{table}

\begin{table}
\renewcommand{\arraystretch}{1.5}
\begin{tabular}{|c|c|c|c| }
\hline
i & x$_{i}$ & y$_{i}$ & z$_{i}$ \tabularnewline
\hline
1 & $\sqrt{\frac{1}{15}}$ & $\sqrt{\frac{1}{15}}$ & $\sqrt{\frac{13}{15}}$\tabularnewline
2 & $\sqrt{\frac{1}{15}}$ & $\sqrt{\frac{7}{15}}$ & $\sqrt{\frac{7}{15}}$\tabularnewline
3 & $\sqrt{\frac{1}{15}}$ & $\sqrt{\frac{13}{15}}$ & $\sqrt{\frac{1}{15}}$\tabularnewline
4 & $\sqrt{\frac{7}{15}}$ & $\sqrt{\frac{1}{15}}$ & $\sqrt{\frac{7}{15}}$\tabularnewline
5 & $\sqrt{\frac{7}{15}}$ & $\sqrt{\frac{7}{15}}$ & $\sqrt{\frac{1}{15}}$\tabularnewline
6 & $\sqrt{\frac{13}{15}}$ & $\sqrt{\frac{1}{15}}$ & $\sqrt{\frac{1}{15}}$\tabularnewline
\hline
\end{tabular}
\caption{\quad{}Quadrature formula Carlson A6 for integration over a sphere: ray directions in the first octant.}
\label{tab:A6}
\end{table}

\begin{figure}
\includegraphics[width=\columnwidth]{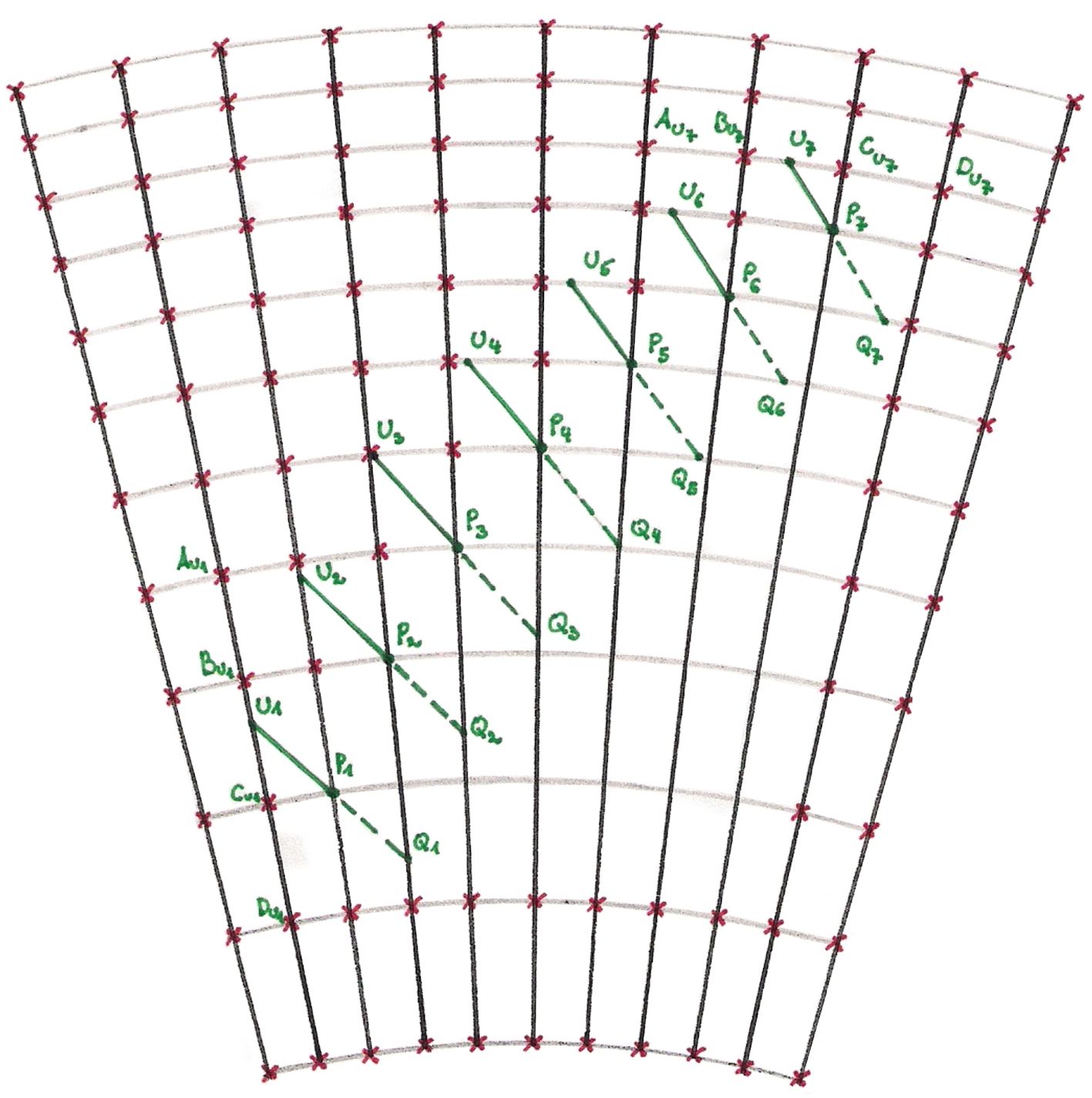}
\caption{\quad{}One step in the iterative solution of the RTE. The rays enter at U. There, at the center P, and
at the exit point Q the opacity has to be determined to perform the numerical integration on the right 
hand side of Eq.~(\ref{RTE}) and to get the intensity $I$ at P for the next step.} \label{fig:rays}       
\end{figure}

For each ray the points of entrance and exit plus the corresponding distances are determined 
(Fig.~\ref{fig:rays}). Since the grid moves this has to be redone at every time step. The one dimensional RTE is solved 
along each ray. 

To determine the incoming intensity along this ray direction one has to interpolate the values of the required
physical quantities to the points U and Q from the neighbouring grid points. This is achieved by linear 
interpolation. Consequently, 
 in two dimensions 4 points are
used for all quantities except for the intensity where 2 points are sufficient. Then we evaluate the equation
\begin{equation}
I\left(\tau_{p}\right)=I\left(\tau_{U}\right)\exp\left(\tau_{u}-\tau_{p}\right)+\int_{\tau_{U}}^{\tau_{P}}S\left(t\right)\exp\left(t-\tau_{p}\right)dt \label{RTE}
\end{equation}
numerically to get $I\left(\tau_{P}\right)$ from a quadrature rule proposed by \cite{kun_olson}. This procedure
is repeated recursively, because after the first step one obtains just the intensity on a single new point.

The mean intensity $J$ is computed by solid angle integration. For A4 the weights are identical,
\begin{equation}
 J=\sum_{i=1}^{N_{\mathrm{rays}}}\frac{1}{N_{\mathrm{rays}}}I\left(r^{i}\right).
\end{equation}
Finally, the radiative heating rate for grey radiative transfer is given by
\begin{equation}
Q_{\mathrm{rad}}=4\pi\rho\chi\left(J-S\right),
\end{equation}
as in the case of a non-moving grid (details on how this is handled within
ANTARES can be found in \cite{muth_na10}).

%
 %
\begin{table}
\begin{tabular}{|c|r|r|r|r|r|r| }
\hline
k & r & j=0 & j=1 & j=2 & j=3 & j=4\tabularnewline
\hline  
1 & -1 & 1 &  &  &  &\tabularnewline
 & 0 & 1 &  &  &  &\tabularnewline
\hline
2 & -1 & 3/2 & -1/2 &  &  & \tabularnewline               
 & 0 & 1/2 & 1/2 &  &  & \tabularnewline                     
 & 1 & -1/2 & 3/2 &  &  & \tabularnewline
\hline
3 & -1 & $15/8$ & -5/4 & 3/8 &  & \tabularnewline      
 & 0 & $3/8$ & 3/4 & -1/8 &  &\tabularnewline
 & 1 & $-1/8$ & 3/4 & 3/8 &  &\tabularnewline
 & 2 & $3/8$ & -5/4 & 15/8 &  & \tabularnewline
\hline 
4 & -1 & $35/16$ & -35/16 & 21/16 & -5/16 &\tabularnewline
 & 0 & $5/16$ & 15/16 & -5/16 & 1/16 &\tabularnewline
 & 1 & $-1/16$ & 9/16 & 9/16 & -1/16 &\tabularnewline
 & 2 & $1/16$ & -5/16 & 15/16 & 5/16 & \tabularnewline 
 & 3 & $-5/16$ & 21/16 & -35/16 & 35/16 & \tabularnewline
\hline 
5 & -1 & $315/128$ & -105/32 & 189/64 & -45/32 & 35/128\tabularnewline
 & 0 & $35/128$ & 35/32 & -35/64 & 7/32 & -5/128 \tabularnewline
 & 1 & $-5/128$ & 15/32 & 45/64 & -5/32 & 3/128\tabularnewline\
 & 2 & $3/128$ & -5/32 & 45/64 & 15/32 & -5/128\tabularnewline 
 & 3 & $-5/128$ & 7/32 & -35/64 & 35/32 & 35/128 \tabularnewline
 & 4 & $35/128$ & -45/32 & 189/64 & -105/32 & 315/128\tabularnewline
\hline 
\end{tabular}\caption{\quad{}The constants $C_{rj}$ for $k=1,\ldots,5$, $j=0,\ldots,4$ on an equidistant grid. }
\label{tab:equidist}
\end{table}

\subsection{WENO schemes on stretched and co-moving polar grids}
\label{subsec:WENO}

The differential equations solved by the ANTARES code can also be written as a system of conservation
laws of the form
\begin{equation}
\partial_{t}U=-\nabla\cdot F\left(U\right)+S.
\end{equation}
Here, $F$ is the vector-valued flux function and $S$ designates (physical or geometrical) source terms. 
A discretisation of the physical domain results in a centre grid $r_{i}$ and a boundary grid $r_{i+\frac{1}{2}}$, 
where the function $F_{i}\left(U\right)$ is computed at the cell centre and interpolated to the cell boundaries.
If a simple polynomial interpolation from the centre grid to the boundary grid is used, shocks or discontinuities 
are either smoothed out or begin to oscillate. To avoid these problems various essentially non-oscillatory (ENO) 
schemes (\cite{harten}, \cite{shu1}, \cite{shu2}, \cite{fed1}) have been implemented into the ANTARES code, where 
interpolation is done by adaptive upwinding stencils. In the ANTARES code the spatial discretisation is done for 
each direction separately. However, note that for a polar grid the radial and angular direction are different. In 
radial direction the grid is stretched while it is equidistant in angular direction. The basic idea of characteristic 
numerical schemes is to transform this nonlinear system into a system of nearly independent \textit{scalar} equations,  
a typical one of the form
\begin{equation}
u_{t}+f(u)_{x}=0
\end{equation}
(possibly with an inhomogeneity at the right hand side) 
and discretise each scalar equation independently and then transform the discretised system back into the 
original variables. Thus, for hyperbolic systems the set of conservation laws is first transformed to the 
eigensystem, where the equations decouple and the upwinding directions can be chosen. For the moving grid  
the eigenvectors are left unchanged as the grid velocity $u_{g}$ only enters into the eigenvalues:
\begin{equation}
\left(\begin{array}{l}
u_{x}-u_{g}-c\\
u_{x}-u_{g}\\
u_{x}-u_{g}\\
u_{x}-u_{g}+c\end{array}\right)
\end{equation}

%
 %
\begin{table*}
 \centering
 \begin{minipage}{130mm}
\renewcommand{\arraystretch}{1.8}
\begin{tabular}{|c|r|c|c|c}
\hline 
k & r & j=0 & j=1 & j=2\tabularnewline
\hline
3&-1&$\frac{a^{6}+2a^{5}+3a^{4}+3a^{3}+3a^{2}+2a+1}{a^{2}\left(a^{4}+2a^{3}+2a^{2}+2a+1\right)}$&$-\frac{a^{4}+a^{3}+a^{2}+a+1}{a^{4}\left(a^{2}+2a+1\right)}$&$\frac{a^{2}+a+1}{a^{4}\left(a^{4}+2a^{3}+2a^{2}+2a+1\right)}$  \tabularnewline
 &0&$\frac{a^{2}\left(a^{2}+a+1\right)}{a^{4}+2a^{3}+2a^{2}+2a+1}$&$\frac{a^{2}+a+1}{a\left(a^{2}+2a+1\right)}$&$-\frac{1}{a\left(a^{4}+2a^{3}+2a^{2}+2a+1\right)}$ \tabularnewline
&1& $-\frac{a^{5}}{a+2a^{3}+2a^{2}+2a+1}$ &$\frac{a\left(a^{2}+a+1\right)}{a+2a+1}$& $\frac{a^{2}+a+1}{a^{4}+2a^{3}+2a^{2}+2a+1}$ \tabularnewline
&2&$\frac{a^{6}\left(a^{2}+a+1\right)}{a^{4}+2a^{3}+2a^{2}+2a+1}$&$-\frac{a^{2}\left(a^{4}+a^{3}+a^{2}+a+1\right)}{a^{2}+2a+1}$&$\frac{a^{6}+2a^{5}+3a^{4}+3a^{3}+3a^{2}+2a+1}{a^{4}+2a^{3}+2a^{2}+2a+1}$ \tabularnewline
\hline
\end{tabular}
\end{minipage}
\caption{\quad{}The constants $C_{rj}$ for $k=3$, $j=0,\ldots,2$ on the stretched grid. For better readability, $a\equiv\sqrt{q}$.}
\label{tab:k=3}
\end{table*}

%
 %
\begin{table*}
 \centering
 \begin{minipage}{120mm}
\renewcommand{\arraystretch}{1.8}
\begin{tabular}{c|c|c}
\hline
j &k=5, r=2 \tabularnewline
\hline
0 & $\frac{a^{16}\left(a^{2}+a+1\right)}{a^{14}+3a^{13}+5a^{12}+8a^{11}+11a^{10}+13a^{9}+15a^{8}+16a^{7}+15a^{6}+13a^{5}+11a^{4}+8a^{3}+5a^{2}+3a+1}$  \tabularnewline
1 & $-\frac{a^{8}\left(a^{4}+a^{3}+a^{2}+a+1\right)}{ a^{8}+3a^{7}+4a^{6}+5a^{5}+6a^{4}+5a^{3}+4a^{2}+3a+1}$ \tabularnewline
2 & $\frac{a^{2}\left(a^{8}+3a^{7}+6a^{6}+8a^{5}+9a^{4}+8a^{3}+6a^{2}+3a+1\right)}{ a^{8}+4a^{7}+8a^{6}+12a^{5}+14a^{4}+12a^{3}+8a^{2}+4a+1}$ \tabularnewline
3 & $\frac{a^{6}+2a^{5}+3a^{4}+3a^{3}+3a^{2}+2a+1}{a\left(a^{8}+3a^{7}+4a^{6}+5a^{5}+6a^{4}+5a^{3}+4a^{2}+3a+1\right)}$ \tabularnewline
4 & $-\frac{a^{4}+a^{3}+a^{2}+a+1}{a\left(a^{14}+3a^{13}+5a^{12}+8a^{11}+11a^{10}+13a^{9}+15a^{8}+16a^{7}+15a^{6}+13a^{5}+11a^{4}+8a^{3}+5a^{2}+3a+1\right)}$ \tabularnewline
\hline
\end{tabular}
\end{minipage}
\caption{\quad{}The constants $C_{2j}$ for $k=5$ and $r=2$.}
\label{tab:k=5}
\end{table*}

Schemes of the ENO variety as first introduced by Harten, Engquist and Chakravarthy (\cite{harten}) 
are based on cell 
averages as follows. Given the cell averages $\overline{f}_{i}\equiv f\left(r_{i}\right)$ of a function we want to find a numerical flux function $\hat{f}_{i+\frac{1}{2}}\equiv\hat{f}\left(f_{i-r},\ldots,f_{i+s}\right)$ such that the flux difference approximates the derivative $f'\left(r_{i}\right)$ where $f$ is sufficiently smooth. The mapping from the given cell averages $\left\{ \overline{f}_{j}\right\}$ in the stencil $S\left(i\right)$ to the values $\hat{f}_{i+1/2}^{-}$ is linear. Therefore, there exist constants $c_{rj}$ and $\tilde{c}_{rj}$ depending on the left shift \textit{r} of the stencil $S\left(i\right)$ such that
\begin{equation}
\hat{f}_{i+1/2}^{-}  = \sum_{j=0}^{k-1}c_{rj}\overline{f}_{i-r+j}
\end{equation}
and
\begin{equation}
\hat{f}_{i-1/2}^{+}  = \sum_{j=0}^{k-1}\tilde{c}_{rj}\overline{f}_{i-r+j}
\end{equation}
where $\tilde{c}_{rj}=c_{r-1,j}$.
Near discontinuities in the solution of hyperbolic conservation laws oscillations can occur because the stencil 
contains those discontinuities. Therefore, an adaptive stencil $S\left(i\right)$ is chosen for the interpolation of the cell 
boundary fluxes where the left shift \textit{r} changes with the location $r_{i}$. After all the main idea of the 
ENO approximation is to exclude cells containing discontinuities from the stencil $S\left(i\right)$.

When using WENO (\cite{liu1}, \cite{jiang}) instead of choosing a single stencil for the interpolation polynomial in 
the ENO reconstruction, a convex combination of all candidates is used to achieve the essentially non-oscillatory 
property. A weighted combination of adaptive stencils gives high order polynomial approximations of the 
divergence term $\nabla\cdot F\left(U\right)$ in smooth regions. Across discontinuities the smoothest stencil is
chosen even though it is of lower order. Instead of performing a $2k-1$ order ENO scheme using stencils of 
that order a combination of $k$ stencils of order $k$ is used to obtain the final accuracy $2k-1$. For ANTARES 
the orders 3 and 5 have been considered. For a \textit{k}-th order ENO scheme there are $k$ candidate stencils
\begin{eqnarray}
S_{r}\left(i\right)=\left\{ x_{i-r},\ldots,x_{i-r+k-1}\right\} ,\quad r=0,\ldots,k-1
\end{eqnarray}
which produce $k$ different reconstructions of the value $f_{i+\frac{1}{2}}$:
\begin{eqnarray}
\hat{f}_{i+1/2}^{\left(r\right)}=\sum_{j=0}^{k-1}c_{rj}\overline{f}_{i-r+j},\quad r=0,\ldots,k-1.
\end{eqnarray}
The convex combination of the values $\hat{f}_{i+1/2}^{\left(r\right)}$
for the WENO approach is 
\begin{eqnarray}   \label{eq:wrweno}
\hat{f}_{i+1/2}=\sum_{r=0}^{k-1}w_{r}\hat{f}_{i+1/2}^{\left(r\right)},\quad r=0,\ldots,k-1
\end{eqnarray}
and is used as a new approximation for $\bar{f}_{i+1/2}$. For the weights $w_{r}\geq 0$ for all $r$
and $\sum_{r=0}^{k-1}w_{r}=1$ must be true. For a smooth function $f\left(x\right)$ there are constants
$d_{r}$ so that $\sum_{r=0}^{k-1}d_{r}=1$ and
\begin{equation}
\hat{f}_{i+1/2}^{-}=\sum_{r=0}^{k-1}d_{r}\hat{f}_{i+1/2}^{\left(r\right)}=f\left(r_{i+1/2}\right)+O\left(\triangle r_{i}^{2k-1}\right)
\label{eq:eno7}
\end{equation}
and constants $\widetilde{d}_{r}$ so that $\sum_{k=0}^{k-1}\tilde{d}_{r}=1$ and
\begin{equation}
\hat{f}_{i-1/2}^{+}=\sum_{r=0}^{k-1}\tilde{d}_{r}\hat{f}_{i+1/2}^{\left(r\right)}=f\left(r_{i-1/2}\right)+O\left(\triangle r_{i}^{2k-1}\right).
\label{eq:eno8}
\end{equation}
In \cite{liu1} weights of the form
\begin{equation}
w_{r}=\frac{\alpha_{r}}{\sum_{s=0}^{k-1}\alpha_{s}}\label{eq:eno6}
\end{equation}
with $\alpha_{r}=\frac{d_{r}}{\left(\varepsilon+\beta_{r}\right)^{2}}$
are proposed. $\varepsilon>0$ is the machine accuracy, which is introduced
here to ensure that the denominator does not become zero. $\beta_{r}$ are
smoothness indicators of the stencil supposed to be zero when a discontinuity
is contained in the stencil. A robust choice of smoothness indicators is defined
by 
\begin{equation}  \label{eq:betaformula}
\beta_{r}=\sum_{l=1}^{k-1}\intop_{r_{i-1/2}}^{r_{i+1/2}}\Delta r_{i}^{2l-1}\left(\frac{\partial^{l}p_{r}\left(x\right)}{\partial x^{l}}\right)dx\end{equation}
where $p_{r}\left(x\right)$ is the reconstruction polynomial on the stencil $S_{r}\left(i\right)$. The smoothness indicators $\beta_{r}$
are a measure for the total variation in the interval $I_{i}$.

From a different point of view another set of approximation coefficients $C_{rj}$ 
(cf.\ Tables \ref{tab:equidist}--\ref{tab:k=5}), of weights $w_{r}$ (cf.\ Table~\ref{D}), of their approximations $D_{r}$ and $\tilde{D}_{r}$ and of
smoothness indicators, denoted by being capitalised, can be derived. In this case the smoothness indicators for $k=3$ are:
\begin{equation} \label{eq:betaformula1}
\begin{aligned}
\beta_{0} & = & Y_{0}^{2}\left(\bar{f}_{i}-\left(q+1\right)\bar{f}_{i+1}+\bar{f}_{i+2}\right)^{2}+\\
&&\left(Z_{10}\bar{f}_{i}+Z_{20}\bar{f}_{i+1}+Z_{30}\bar{f}_{i+2}\right)^{2}\\
\beta_{1} & = & Y_{1}^{2}\left(\bar{f}_{i}-\left(q+1\right)\bar{f}_{i+1}+\bar{f}_{i+2}\right)^{2}+\\
&&\left(Z_{11}\bar{f}_{i-1}+Z_{21}\bar{f}_{i}+Z_{31}\bar{f}_{i+1}\right)^{2}\\
\beta_{2} & = & Y_{2}^{2}\left(\bar{f}_{i}-\left(q+1\right)\bar{f}_{i+1}+\bar{f}_{i+2}\right)^{2}+\\
&&\left(Z_{12}\bar{f}_{i-2}+Z_{22}\bar{f}_{i-1}+Z_{32}\bar{f}_{i}\right)^{2}
\end{aligned}
\end{equation}
where $Y$ and $Z_{i}$ are given in Table~\ref{beta}. These are also available in the ANTARES code in addition
to those from (\ref{eq:betaformula}) which are based on the approach (\ref{eq:wrweno})--(\ref{eq:eno6})
and are required for stable time integrations of our Cepheid models, as we show below. In the second approach 
instead of viewing $f\left(r_{i}\right)$ as cell averages and interpolating the primitive function, the flux function
$f\left(r_{i}\right)$ itself is interpolated to the boundaries, leading to a different set of coefficients
(Table~\ref{tab:equidist}). In our case a combination of stencils of length 3 with coefficients $C_{rj}$ from 
Table~\ref{tab:k=3} yields the stencil of length 5 with coefficients  $C_{2j}$ in Table~\ref{tab:k=5} in smooth 
regions. This approach was chosen because we use a polar grid and although the spacing is that of a stretched 
grid in radial direction, the cell-volumes depend also on the radius as does the shape of the cells. 
 
\begin{figure}
\includegraphics[width=\columnwidth]{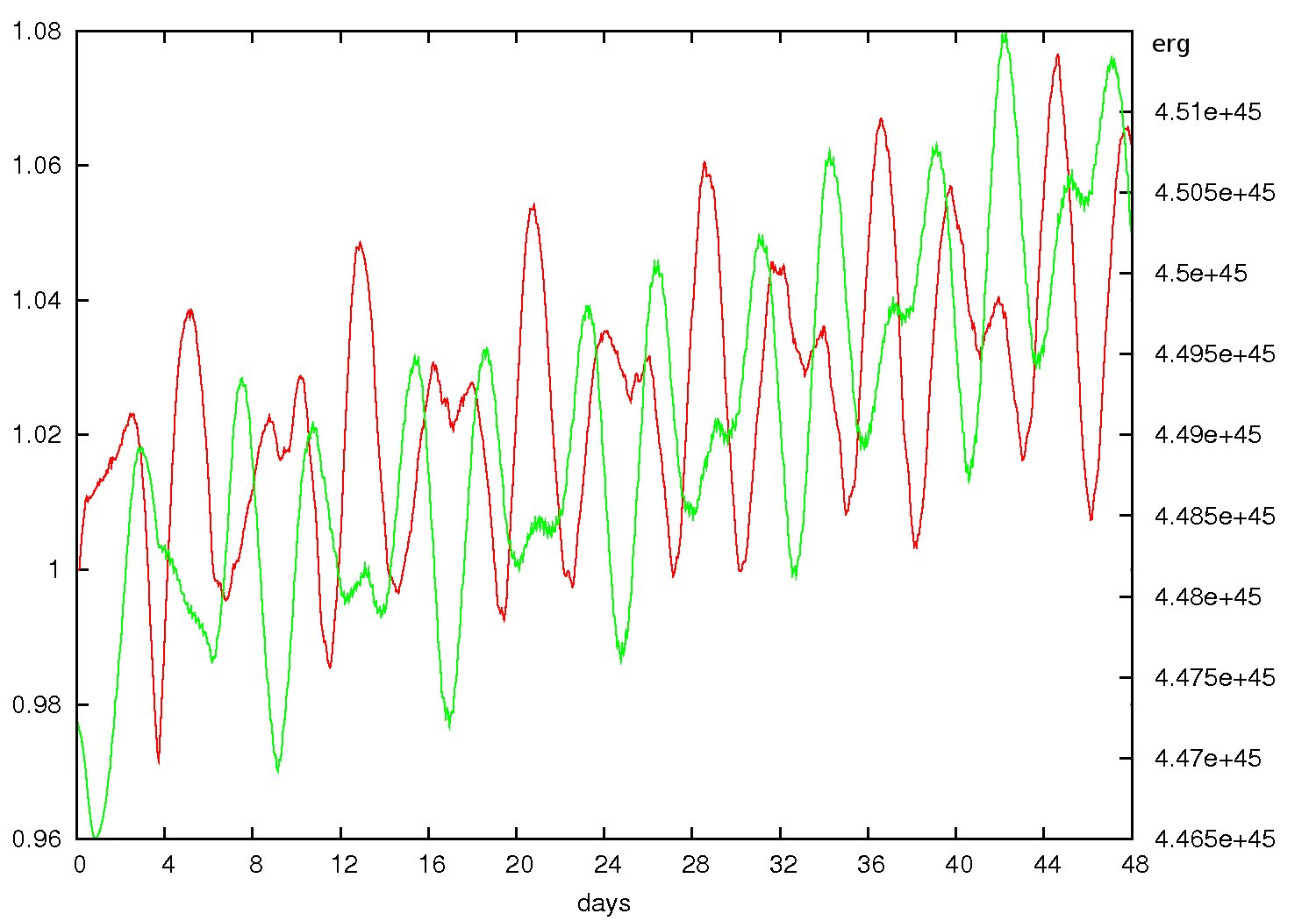}
\caption{\quad{} Cepheid simulation with interpolation of the primitive function with cell averages in all directions: 
time evolution of radius (red) and total energy (green). Abscissa: time in days, left ordinate: radius on a linear scale
relative to its initial value, right ordinate: total energy within the sphere in erg.}
\label{fig:shucoeffs}       
\end{figure}

\begin{figure}
\includegraphics[width=\columnwidth]{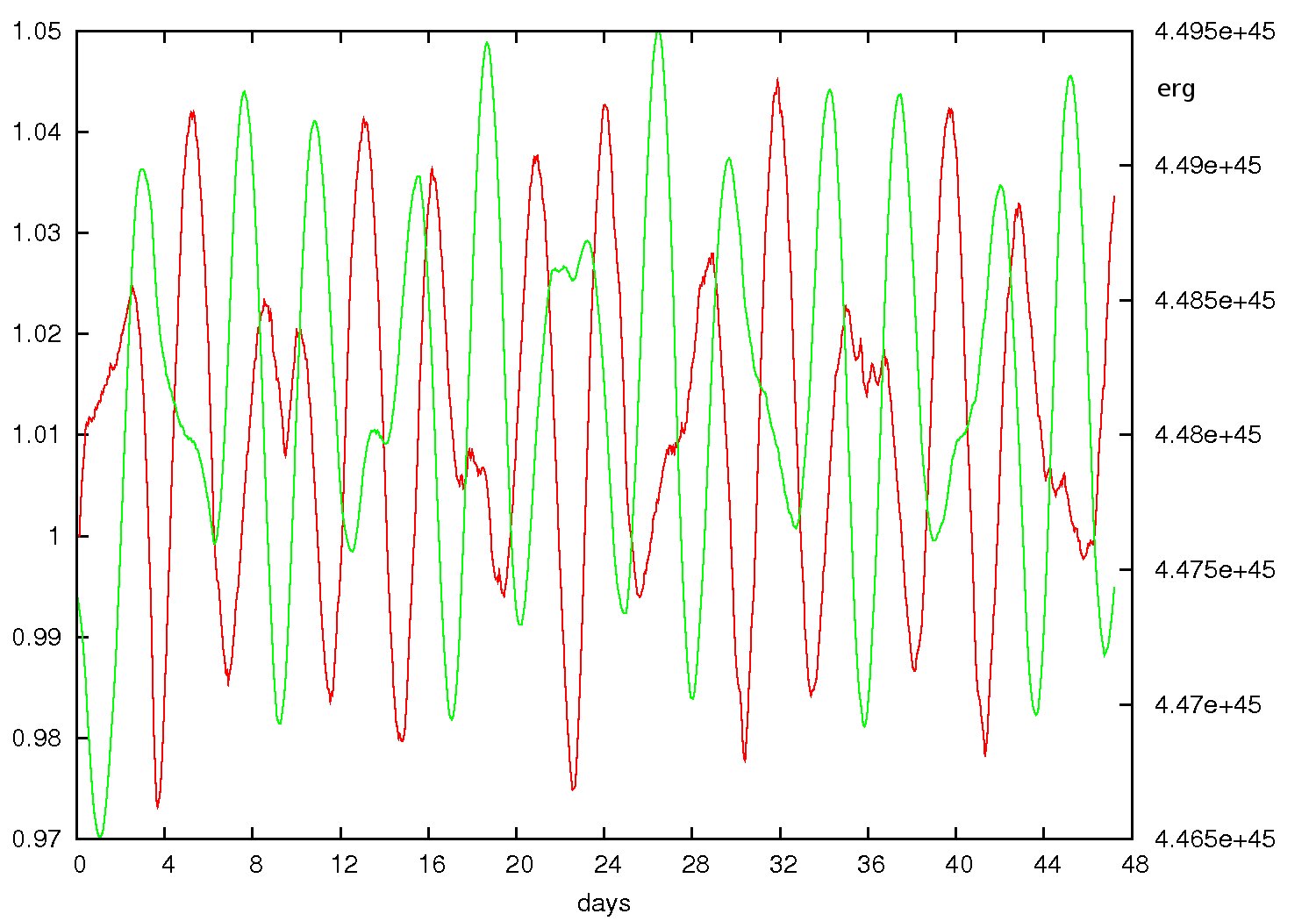}
\caption{\quad{} Cepheid simulation with interpolation of the flux function and the new scheme used in all directions: 
time evolution of radius (red) and total energy (green). Abscissa: time in days, left ordinate: radius on a linear 
scale relative to its initial value, right ordinate: total energy within the sphere in erg.}
\label{fig:antcoeffs}       
\end{figure}

The approach via exact cell volumes (\ref{eq:wrweno})--(\ref{eq:betaformula}) is not advisable in 
our case since we want to ensure conservation of the basic variables. 
Though the difference in area size between a rectangular cell and a polar cell is very small the difference
over the whole grid is not and the ensuing error cannot be neglected. When using the original coefficients in 
a one-dim\-ensional simulation e.g. energy is not preserved over a prolonged period of time: during the first 48 days of a Cepheid 
simulation it increases or decreases depending on various conditions. (The parameters for this 
Cepheid are provided Sect.~\ref{subsec:modelparameters}). One run shows increasing values (see Fig.~\ref{fig:shucoeffs}),  
but for a slightly larger domain and with artificial diffusivities included we obtain decreasing values. 
With the new set of coefficients all the conserved variables are stable (Fig.~\ref{fig:antcoeffs}) when averaged
over several pulsation cycles. This comparison also demonstrates that a non-conservative discretisation method is unacceptable for a realistic 
numerical simulation of Cepheids.

\begin{table}
\renewcommand{\arraystretch}{1.8}
\begin{tabular}{l|c}
\hline 
$D_{0}$ & $\frac{a^{4}+a^{3}+a^{2}+a+1}{a^{10}+a^{9}+a^{8}+2a^{7}+2a^{6}+2a^{5}+2a^{4}+2a^{3}+a^{2}+a+1}$\tabularnewline 
$D_{1}$ & $\frac{a^{3}\left(a^{4}+a^{3}+a^{2}+a+1\right)}{a^{8}+a^{7}+a^{5}+2a^{4}+a^{3}+a+1}$\tabularnewline
$D_{2}$ & $\frac{a^{10}}{a^{10}+a^{9}+a^{8}+2a^{7}+2a^{6}+2a^{5}+2a^{4}+2a^{3}+a^{2}+a+1}$\tabularnewline
\hline
$\tilde{D}_{0}$ & $\frac{1}{a^{10}+a^{9}+a^{8}+2a^{7}+2a^{6}+2a^{5}+2a^{4}+2a^{3}+a^{2}+a+1}$\tabularnewline 
$\tilde{D}_{1}$ & $\frac{a\left(a^{4}+a^{3}+a^{2}+a+1\right)}{a^{8}+a^{7}+a^{5}+2a^{4}+a^{3}+a+1}$\tabularnewline
$\tilde{D}_{2}$ & $\frac{a^{6}\left(a^{4}+a^{3}+a^{2}+a+1\right)}{a^{10}+a^{9}+a^{8}+2a^{7}+2a^{6}+2a^{5}+2a^{4}+2a^{3}+a^{2}+a+1}$\tabularnewline
\hline
\end{tabular}\caption{\quad{}Weights $D_{r}$ for $\hat{f}_{i+1/2}^{-}$ and $\tilde{D}_{r}$ for $\hat{f}_{i-1/2}^{+}$ for $k=3$.}
\label{D}
\end{table}

\begin{table}
\begin{tabular}{|c|c|c|c|c}
\hline 
j  & $Y_{j}$ & $Z_{1j}$ &$Z_{2j}$ & $Z_{3j}$\tabularnewline
\hline 
$ 0 $ &$\frac{\sqrt{13}}{\sqrt{3}a^{4}\left(a^{2}+1\right)}$ & $\frac{a^{2}+a+1}{a^{2}\left(a+1\right)}$ & $-\frac{a^{4}+a^{3}+a^{2}+a+1}{a^{4}\left(a+1\right)}$ & $\frac{1}{a^{4}\left(a+1\right)}$\tabularnewline
$ 1 $ &$\frac{\sqrt{13}}{\sqrt{3}\left(a^{2}+1\right)}$ & $\frac{a^{2}}{a+1}$ & $\frac{-a^{3}+1}{a\left(a+1\right)}$ & $-\frac{1}{a\left(a+1\right)}$\tabularnewline
$ 2 $ & $\frac{\sqrt{13}a^{4}}{\sqrt{3}\left(a^{2}+1\right)}$ & $\frac{a^{5}}{a+1}$ & $-\frac{a\left(a^{4}+a^{2}+a+1\right)}{a+1}$ & $\frac{a\left(a^{2}+a+1\right)}{a+1}$\tabularnewline
\hline
\end{tabular}\caption{\quad{}Coefficients for evaluating $\beta_{j}$, $j=0,1,2$, for $k=3$, Equation~\ref{eq:betaformula1}.}
\label{beta}
\end{table}


Note that the formally fifth order of the WENO5 scheme \citep{shuGal} does no longer hold with the new 
approach. However, lack of energy conservation is not an acceptable alternative for locally higher order
in smooth regions. Around discontinuities both old and new coefficient sets yield methods of
second order which is also the spatial order of accuracy achieved with the new set in smooth
regions. Moreover, for the one-dimensional linear advection equation with constant velocity 
the error coefficient is one quarter of the size of standard second order methods such as the
leap-frog scheme. Leap-frog time integration, however, can cope with shocks only after
modifications that degrade their accuracy even further. The main computational expense 
of the WENO approach is the calculation of the fluxes at each cell boundary. This occurs only
once, whence the overhead due to larger interpolation stencils is negligible, as they merely
use already computed information. The only important restriction originating from wider
stencils in comparison with traditional second order shock capturing methods
are somewhat larger minimum domain sizes for the domain decomposition
approach used in the MPI parallelization in ANTARES.

\subsection{Discretisation scheme for diffusion}   \label{subsec:diffusion}

\begin{table}
\renewcommand{\arraystretch}{1.7}
\setlength{\tabcolsep}{1pt}
\begin{tabular}{r|c|r}
\hline
&&$q=1$\tabularnewline
\hline
-2&$ -\frac{3+6a+9a^{2}+10a^{3}+11a^{4}+10a^{5}+9a^{6}+6a^{7}+4a^{8}+2a^{9}+a^{10}}{a^{2}\left(1+a\right)^{2}\left(1+2a^{2}+2a^{4}+a^{6}\right)}$& $-\frac{71}{24}$ \tabularnewline
-1&$  \frac{3+4a+5a^{2}+6a^{3}+7a^{4}+6a^{5}+6a^{6}+4a^{7}+3a^{8}+2a^{9}+a^{10}}{a^{6} \left(1+a\right)^{2}\left(1+a^{2}\right)}$& $\frac{47}{8}$\tabularnewline
0&$  -\frac{3+4a+5a^{2}+4a^{3}+4a^{4}+4a^{5}+4a^{6}+2a^{7}+a^{8}}{a^{8} \left(1+a\right)^{2}\left(1+a^{2}\right)}$& $-\frac{31}{8}$\tabularnewline
1  &$\frac{3+4a+5a^{2}+4a^{3}+4a^{4}+2a^{5}+a^{6}}{a^{8}\left(1+a\right)^{2}\left(1+2a^{2}+2a^{4}+a^{6}\right)}$& $\frac{23}{24}$\tabularnewline
\hline
-2&$  -\frac{3a^{4}+4a^{5}+5a^{6}+4a^{7}+4a^{8}+2a^{9}+a^{10}}{\left(1+a\right)^{2}\left(1+2a^{2}+2a^{4}+a^{6}\right)}$& $-\frac{23}{24}$ \tabularnewline
-1&$ \frac{-2a-a^{2}+2a^{4}+2a^{5}+3a^{6}+2a^{7}+a^{8}}{a^{2}\left(1+a\right)^{2}\left(1+a^{2}\right)}$& $\frac{7}{8}$ \tabularnewline
0&$  \frac{2a-a^{6} }{a^{4}\left(1+a\right)^{2}\left(1+a^{2}\right)}$& $\frac{1}{8}$ \tabularnewline
1  &$ -\frac{2a-a^{4}}{a^{4}\left(1+a\right)^{2}\left(1+2a^{2}+2a^{4}+a^{6}\right)}$ & $-\frac{1}{24}$\tabularnewline
\hline
-2&$\frac{2a^{9}-a^{12}}{\left(1+a\right)^{2}\left(1+2a^{2}+2a^{4}+a^{6}\right)}$& $\frac{1}{24}$\tabularnewline
-1&$-\frac{\left(2a^{3} - a^{4}\right)\left(1+a+a^{2}\right)^{2}}{\left(1+a\right)^{2}\left(1+a^{2}\right)}$& $-\frac{9}{8}$\tabularnewline
0&$\frac{\left(-1+2a\right)\left(1+a+a^{2}\right)^{2}}{\left(1+a\right)^{2}\left(1+a^{2}\right)}$& $\frac{9}{8}$\tabularnewline
1  &$-\frac{\left(-1+2a^{3}\right)}{\left(1+a\right)^{2}\left(1+2a^{2}+2a^{4}+a^{6}\right)}$& $-\frac{1}{24}$\tabularnewline
\hline
-2&$\frac{- a^{12}+2a^{15}}{ (1+a)^{2}(1+2a^{2}+2a^{4}+a^{6})}$& $\frac{1}{24}$\tabularnewline
-1&$-\frac{-a^{6}+2a^{11}}{(1+a)^{2}(1+a^{2})}$& $-\frac{1}{8}$\tabularnewline
 0&$-\frac{a^{2}+2a^{3}+3a^{4}+2a^{5}+2a^{6}-a^{8}-2a^{9}}{(1+a)^{2}(1+a^{2})}$& $-\frac{7}{8}$\tabularnewline
1  &$ \frac{a^{2}+2a^{3}+4a^{4}+4a^{5}+5a^{6}+4a^{7}+3a^{8}}{ (1+a)^{2}(1+2a^{2}+2a^{4}+a^{6})}$& $\frac{23}{24}$\tabularnewline
\hline
-2&$-\frac{a^{14}+2a^{15}+4a^{16}+4a^{17}+5a^{18}+4a^{19}+3a^{20}}{ (1+a)^{2}(1+2a^{2}+2a^{4}+a^{6})}$& $-\frac{23}{24}$\tabularnewline
 -1&$ \frac{a^{8}+2a^{9}+4a^{10}+4a^{11}+4a^{12}+4a^{13}+5a^{14}+4a^{15}+3a^{16}}{(1+a)^{2}(1+a^{2})}$& $\frac{31}{8}$\tabularnewline
0&$-\frac{a^{4}+3a^{6}+6a^{8}+7a^{10}+5a^{12}+3a^{14}+2a^{5}+4a^{7}+6a^{9}+6a^{11}+4a^{13}}{(1+a)^{2}(1+a^{2})}$& $-\frac{47}{8}$\tabularnewline
1  &$ \frac{a^{4}+2a^{5}+4a^{6}+6a^{7}+9a^{8}+10a^{9}+11a^{10}+10a^{11}+9a^{12}+6a^{13}+3a^{14}}{ (1+a)^{2}(1+2a^{2}+2a^{4}+a^{6})}$& $\frac{71}{24}$\tabularnewline
\hline
\end{tabular}
\caption{\quad{}Discretisation of the derivative terms representing diffusion. Weights for the stencil
$\left(i-2,i-1,i,i+1\right)$ for the derivatives at, from top to bottom, $i-2.5$, $i-1.5$, $i-0.5$, $i+0.5$, and $i+1.5$. The 
derivatives at $i-2.5$, $i-1.5$, $i+0.5$, and $i+1.5$ are only needed near the boundaries, in the inner regions
the weights for $i-0.5$ are used (cf.\ Eq.~(\ref{eq_example})). For the equidistant grid these terms reduce to the values
in the rightmost column.  For better readability, $a\equiv\sqrt{q}$.}
\label{tab:diff}
\end{table}

%

To avoid problems with conservation of mass, momentum, and energy as exemplified 
through Fig.~\ref{fig:shucoeffs} the interpolation of the diffusive fluxes has to be
done in the same manner. The new WENO coefficients are given by Tables~\ref{tab:k=3}--\ref{beta}.
\cite{happ11} and \cite{koch10} discussed a differencing scheme for diffusive fluxes
based on interpolating cell averages. For constant diffusivities and equidistantly spaced
grid points in one dimension in a Cartesian coordinate system it recovers 
the well-known approximation
\[
\left(-U_{i}\right)_{xx}=\frac{U_{i-2}-16\, U_{i-1}+30\, U_{i}-16\, U_{i+1}+U_{i+2}}{12\, h^2}
\]
which is of fourth order. However, this cannot be carried over to the case of a stretched, co-moving, 
polar grid for the same reasons as already discussed for the WENO scheme. 
Plain staggered mesh interpolation of fluxes can be used instead of interpolating
the cell averages in which case the derivative at $i+a-\frac{1}{2}$ is computed as
\begin{equation}
\left(U_{i+a-\frac{1}{2}}\right)_{r}=\frac{S_{-2}U_{i-2}+S_{-1}U_{i-1}+S_{0}U_{i}+S_{1}U_{i+1}}{r_{i+1}-r_{i}}
\end{equation}
for $a=-2,...,2$.
For the equidistant grid this leads to 
\begin{equation}  \label{eq_example}
\left(U_{i-\frac{1}{2}}\right)_{r}=\frac{U_{i-2}-27\, U_{i-1}+27\, U_{i}-U_{i+1}}{24\, h}
\end{equation}
and
\begin{equation}
\left(U_{i+\frac{1}{2}}\right)_{r}=\frac{U_{i-1}-27\, U_{i}+27\, U_{i+1}-U_{i+2}}{24\, h}
\end{equation}
and the second derivative is then given by
\begin{equation}
\left(-U_{i}\right)_{rr}=\frac{U_{i-2}-28\, U_{i-1}+54\, U_{i}-28\, U_{i+1}+U_{i+2}}{24\, h^2}
\end{equation}
(the latter was also discussed in \cite{koch10} for Cartesian grids).
This is a second order approximation for $\left(U_{i}\right)_{rr}$ with 
an error coefficient half the size of that one of the central second order difference on
an equidistant grid in one dimension for the heat equation with constant
diffusivity. As with the WENO scheme, the expensive part in evaluating these stencils
is the computation of the cell boundary fluxes, whence the extra accuracy comes at
negligible costs. Moreover, the stencils are still smaller than their WENO counterparts and
thus introduce only small overheads when used in ANTARES in parallel mode. This approach
can be generalised to the stretched and co-moving polar grid in a conservative manner. 
Details are given in Table~\ref{tab:diff} which displays these new stencils for both the interior 
of the domain (third panel) and for points near the (vertical) boundaries.

\subsection{Subgrid scale modelling}
\label{subgrid}

In some of our simulations the grid refinement was aided or substituted by subgrid scale modelling. The approach
we use for this purpose is based on the work of \cite{smagorinsky} and \cite{lilly}. The subgrid-scale stress is written
as
\begin{equation}
\tau_{ij}^{\mathrm{a}}=\tau_{ij}-\frac{1}{3}\delta_{ij}\tau_{kk}=-2\rho K_{m}\tilde{D}_{ij}
\end{equation}
where the superscript ``a'' denotes the anisotropic part of a tensor and $\tilde{D}_{ij}$ is the strain rate tensor,
\begin{equation}
\tilde{D}_{ij}=\frac{1}{2}\left(\frac{\partial\tilde{u}_{i}}{\partial x_{j}}+\frac{\partial\tilde{u}_{j}}{\partial x_{i}}\right)-\frac{1}{3}\delta_{ij}\frac{\partial\tilde{u}_{k}}{\partial x_{k}}.
\end{equation}
The eddy viscosity $K_{m}$ is given by 
\begin{equation}
K_{m}=C\Delta^{2}\left|\tilde{D}\right|,
\end{equation}
where $\Delta$ is the filter width, approximated in 2D as $\left(\Delta r_{i}\Delta y_{i}\right)^{1/2}$,
and as $\left(\Delta r_{i}\Delta y_{i}\Delta z_{i}\right)^{1/3}$ in
three dimensions, and $C$ is a dimensionless coefficient. This
coefficient is a constant parameter which is commonly expressed through the Smagorinsky
coefficient $c_{s}=C^{1/2}$. For our simulations we have chosen $c_{s}=0.2$
which is the standard value found in the literature for 2D simulations
(for 3D simulations a smaller value of about 0.1 is usually preferred).

\subsection{Initial and boundary conditions}
\label{initbound}
We start with a one-dimensional model. For the simulations presented here such a model was kindly provided
by G\"unter Houdek. There is no turbulent pressure included, thus the starting model set up in ANTARES is purely 
radiative. To accelerate the onset of pulsations the gravitational acceleration is reduced by 1\% during the first few
time steps. After the pulsation becomes stable and remains so for a considerable time the one-dimensional model 
is converted into a two-dimensional one by copying the one-dimensional model repeatedly in azimuthal direction.
Slight random perturbations are applied to the new lateral momentum variable to start the 
truly 2D flow.

For closed boundary conditions the density is set to hydrostatic equilibrium at the top. 
Moreover, the conditions
\begin{equation}
\begin{aligned}
u_{r}|_{\mathrm{top}} & = & u_{g}|_{\mathrm{top}}\\
\frac{\partial u_{\varphi}}{\partial r}|_{\mathrm{top}} & = & 0\\
\frac{\partial T}{\partial r}|_{\mathrm{top}} & = & 0
\end{aligned}
\label{equ:closed_top}
\end{equation}
are applied at the top, and 
\begin{equation}
u_{r}|_{\mathrm{bot}}=0
\end{equation}
at the bottom. Since the star is purely radiative at the bottom the
incoming energy transport is adjusted by keeping the radial component
of radiative flux density $\kappa\nabla T$ at its initial value.
After applying the boundary conditions, the grid is updated by moving the grid coordinates with grid velocity $\vec{u}_{g}^{n}$. The new grid velocity $\vec{u}_{g}^{n+1}$is determined as the horizontal average of the vertical velocity at the top.

\section{Computational challenges}
\label{sec:demands}

In this section we discuss, after presenting the parameters of our model,
 various restrictions on the timestep and their interaction with grid refinement. For model~1 the 
timestep varied from 0.12 to 0.16 seconds. This leads to approximately 2.5~million Runge-Kutta steps for the computation 
of one pulsation cycle. The simulation was carried out in MPI-mode on 256 nodes. At continuous operation this allows
the calculation of  one pulsation cycle every five days. For the better resolved models the computational demands were 
even higher.

\subsection{The model parameters for the Cepheid simulations}
\label{subsec:modelparameters}

For our Cepheid models we assume an effective temperature of $T_{\mathrm{eff}}=5125~\mathrm{K}$, 
a surface gravity of $92.48~\mathrm{cm\, s^{-2}}$ (i.e.\ $\log(g) \sim 1.97$), a luminosity $L \sim 913~L_{\odot}$,
 and a mass of $5~M_{\odot}$. For their composition we assume
a subsolar metallicity, i.e.\ a hydrogen mass fraction of $X=0.7$ and a metal mass fraction of $Z=0.01$
(with the composition of $Z$ taken from \citealt{GN_1993}). These values are well within the parameter 
range for which Cepheids can be observed (see \citealt{bono_2000}).
Through a series of tests we found that these model parameters also have the advantage of reducing the 
numerical demands in comparison with higher luminosities and lower metallicities, for example. 

The stellar radius in our models hence is 26.8~Gm (i.e.\ $R \sim 38.5\, R_{\odot}$), of which the outer 
11.3~Gm were modelled. Three 
different aperture angles were used: $1^{\mathrm{o}}$,  $3^{\mathrm{o}}$, and $10^{\mathrm{o}}$. The 
$3^{\mathrm{o}}$-models calculations were also performed with a grid refinement in the 
hydrogen-ionisation zone with refinement factors of 3 in radial direction and 4 in polar direction. 
Further details such as the radial stretching factor between vertically adjacent grid cells (innermost cells
are the longest ones), the total number of grid points in each direction, and whether subgrid scale
modelling is used or not, are given in Table~\ref{tab:models}. Each of these four models develops 
a first overtone pulsation with a period $P$ of about 3.85 days (this is slightly shorter than one would 
expect if the structure of the entire star were contained in the computational domain). 

\subsection{Time stepping}
\label{subsec:timestep}

If we consider the rate at which the cell average of the basic hydrodynamical variables $\rho$, $\vec{I}$, 
and $e$ changes as a function of time, the largest possible timestep is determined by the CFL-condition.
But in practice all our simulations were limited by the radiative timestep.

In the interior regions the diffusion approximation is valid and thus
\begin{equation}
\Delta t_{\mathrm{diff}}\propto \min\left( \frac{3~c_{\rm p}}{16\kappa\sigma T^{3}}\left( \frac{\kappa\rho}{k}\right)^{2} \right).
\label{eq:DeltaDiff}
\end{equation}
As usual, $c_{\rm p}$ stands for the heat capacity at constant pressure and 
$\sigma$ for the Stefan-Boltzmann constant. 
$k$ is essentially the inverse of the relevant, in our case
therefore minimal, grid spacing, which enforces the most stringent time 
step restriction, i.e.
$k=C/\min(\Delta r_{i},\Delta y_{i})$ with typically 
$1 \leq C \leq 2 \pi$. $C$ depends on the method used for spatial discretisation. 
In the code we use $C=1$ and a constant safety factor less than or equal to 1 is applied in
front of the entire expression. The resulting time scale is very small in the optically thin regions. 
On the other hand, the time scale for relaxing a temperature perturbation of arbitrary optical thickness by 
radiation (assuming the radiative transfer equation where appropriate, not just the
diffusion approximation) was shown by \cite{spiegel} to be, in linear approximation of the perturbation, 
\begin{equation}
\Delta t_{\mathrm{rad}}\propto \min \left( \frac{c_{\rm p}}{16\kappa\sigma T^{3}}\left( 1-\frac{\kappa\rho}{k}
\mbox{\rm arccot}\frac{\kappa\rho}{k}\right)^{-1}\right).
\label{eq:DeltaRad}
\end{equation}
Using radiative transfer in optically thin regions (and therefore applying the 
timestep restrictions expressed by Eq.~(\ref{eq:DeltaRad}) rather than 
Eq.~(\ref{eq:DeltaDiff})) results in considerably larger allowed 
numerical timesteps for our Cepheid. Eq.~(\ref{eq:DeltaRad}) leads to a  
minimum for $\Delta t_{\mathrm{rad}}$ in the optically thick regions where, however, 
$\Delta t_{\mathrm{rad}}$ converges to the time scale of radiative diffusion anyway. The timestep 
restrictions for our explicit methods become ever 
more stringent when considering finer grids. Circumventing them requires an implicit method for the time integration 
of the $Q_{\rm rad}$ term which we have not yet implemented.

\subsection{Resolving steep gradients}
\label{subsec:resolving}

\begin{figure}
\includegraphics[width=\columnwidth]{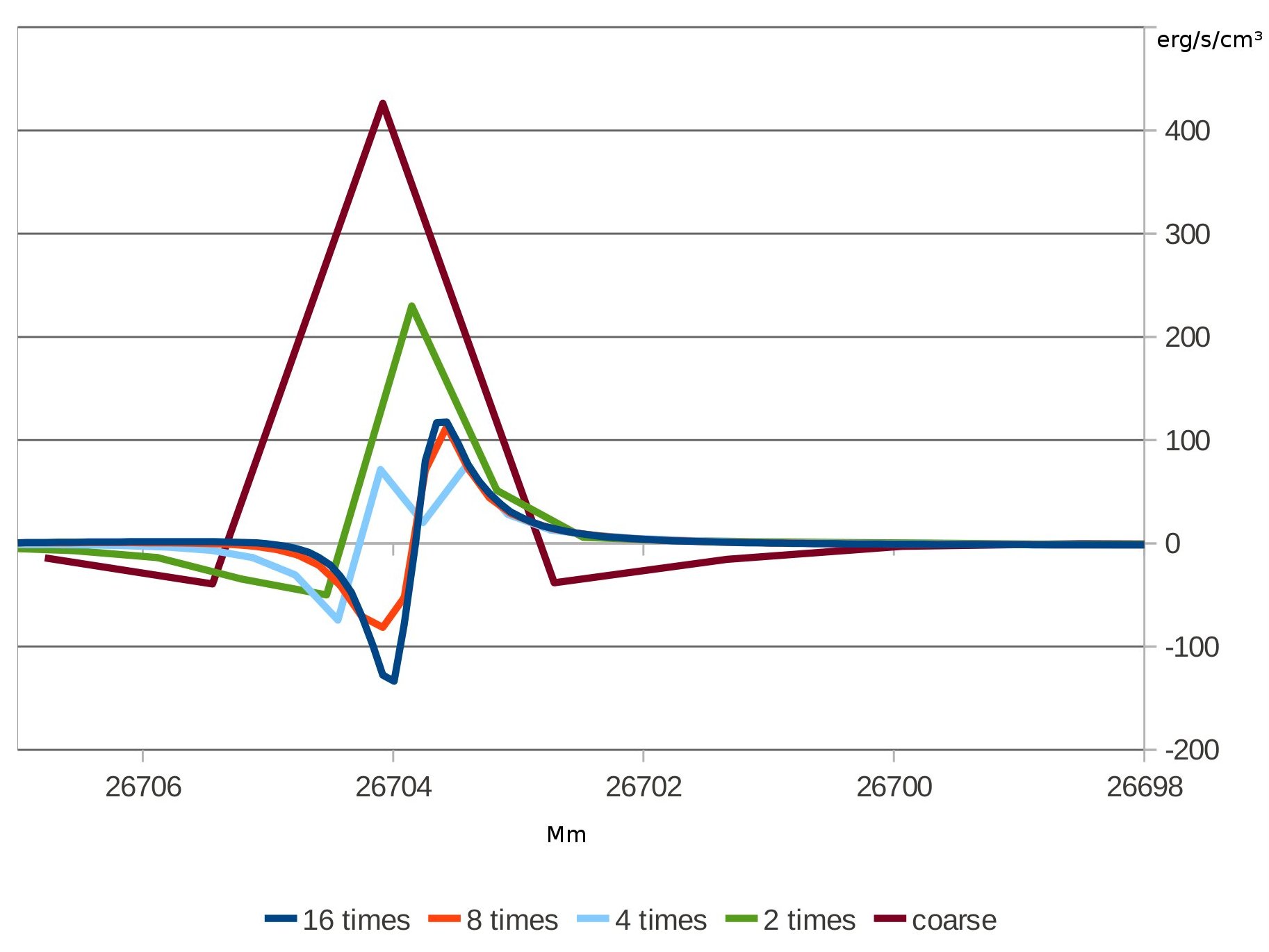}
\caption{\quad{}The shape of $Q_{\mathrm{rad}}$ (given in erg s$^{-1}$ cm$^{-3}$) for a one-dimensional 
simulation at different grid refinement factors.}
\label{fig:qrad}       
\end{figure}

To resolve the steep gradients of density, temperature etc. in the hydrogen ionisation zone a resolution finer  
than 510 points along the radial direction is needed. For a first estimate of the required spacing we compare
the pressure scale heights $H_{p}=p/|\nabla p|$ in the form $p/(\rho g)$ for models of Cepheids and the Sun
in the region of interest, i.e.\ the stellar photosphere at a Rosseland optical depth of order unity. Substituting 
the pressure of a perfect gas, $p_{\mathrm{gas}} = \rho\, \mathrm{R}_{\mathrm{gas}}T / \mu$, for the pressure $p$ 
and the effective temperature $T_{\mathrm{eff}}$ for the temperature $T$ thus yields a ratio of pressure 
scale heights
\begin{equation}  \label{eq:Hp}
\frac{H_{p}}{H_{p\,\odot}}=\frac{T_{\mathrm{eff}}\,\mu_{\odot}\, g_{\odot}}{T_{\mathrm{eff}\,\odot}\,\mu\, g}.
\end{equation}

In our simulations of solar surface convection it was found that a minimum spacing of 20~km in radial 
direction is necessary to just barely resolve all basic thermodynamical variables as a function of depth. 
This translates to approximately 4.25~Mm for our particular Cepheid. But we must also take into account the 
superadiabatic gradient $\nabla-\nabla_{\rm ad}$, because we have to resolve the temperature
gradient equally well as the pressure gradient. This may not be the case when considering only
Eq.~(\ref{eq:Hp}) to determine the spatial resolution of the simulation box, if the temperature gradient
increases much more rapidly than it does at the superadiabatic peak of the solar convection zone. In the 
Sun $\max(\left(\nabla-\nabla_{\rm ad}\right)_{\odot}) \approx 0.6$ (cf.\ \citealt{rosen99}).
For the initial condition of the Cepheid model, which was derived using a mixing-length treatment (MLT)
of convection, the maximum temperature gradient critically depends on the $\alpha$-parameter. In starting
models where no turbulent pressure is included, $\max(\nabla-\nabla_{\rm ad}) \approx 14$, thus the grid 
spacing would reduce to 0.18~Mm or a grid roughly seven times as fine as the grid used in model 1 at that
location. On the other hand, if the starting model is derived with turbulent pressure included, 
$\max(\nabla-\nabla_{\rm ad}) \approx 3.2$, and the grid of this model would only need to be refined
roughly by a factor of two. While the first scenario is the only possible solution for a one dimensional, 
purely radiative model, there is reason to assume that in more dimensions a smaller factor can be 
achieved after convection has set in: the turbulent pressure caused by the flow supports the 
gas pressure in  counteracting gravity and thus allows for lower gradients in gas pressure and temperature.
The additional computation time is of course substantial: a factor of 
7 in grid refinement leads to a factor $\sim20$ for the number of temporal steps required when the timestep is 
limited by $\Delta t_{\mathrm{rad}}$ as given by Eq.~(\ref{eq:DeltaRad}). We have checked that this estimate
holds at least in 1D and in Fig.~\ref{fig:qrad} we present a comparison after a time evolution
of 0.25~sec from the initial, static model.
Obviously in 1D, a grid refinement of a factor of 2 is not enough, with a factor of 4, the radiative heating term starts to converge toward its actual shape and a factor of 8 suffices. The difference to
a factor of 16 presently is not worth the extra computational costs required by such a fine resolution
although it may still be large enough to lead to measurable effects in calculations of stellar
spectra in future work. For our 2D simulations, where easing occurs due to the turbulent effects mentioned above, we applied a grid refinement factor of 3 in radial direction, and additional subgrid scale modelling. 

\section{Grid refinement}
\label{sec:grid}

\begin{figure*}
\centering
\includegraphics[width=\textwidth,height=7.5cm]{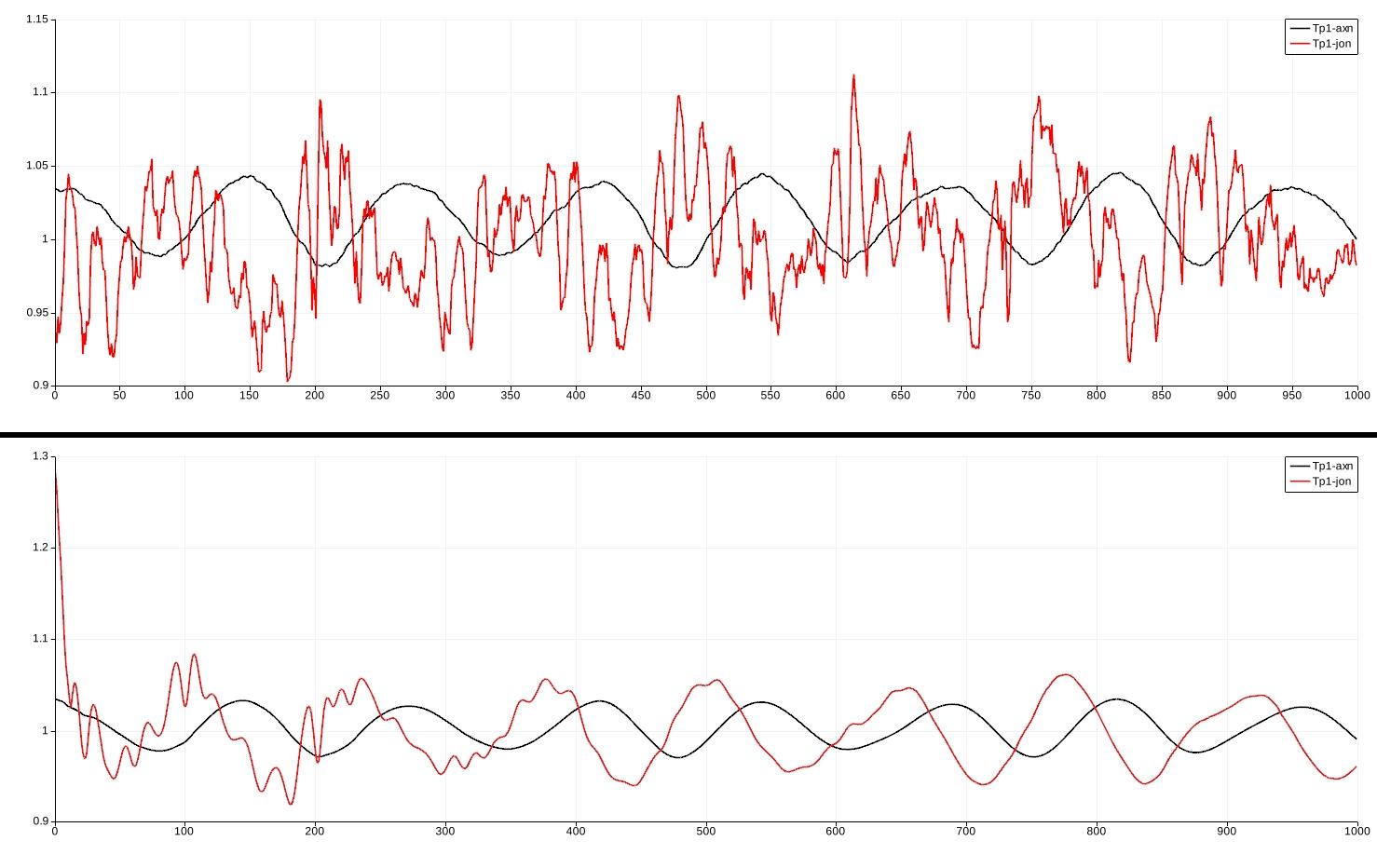}
\caption{\quad{}The light curve for a 1D simulation on the coarse grid and on a grid having a 4 times higher resolution.
              Abscissa: output steps, ordinate: deviation from the average value. In both panels, the red line denotes the
              lightcurve while the black line indicates the radius.}    
\label{fig:1Dfine} 
\end{figure*} 

\begin{figure}
\includegraphics[width=\columnwidth, height=5cm]{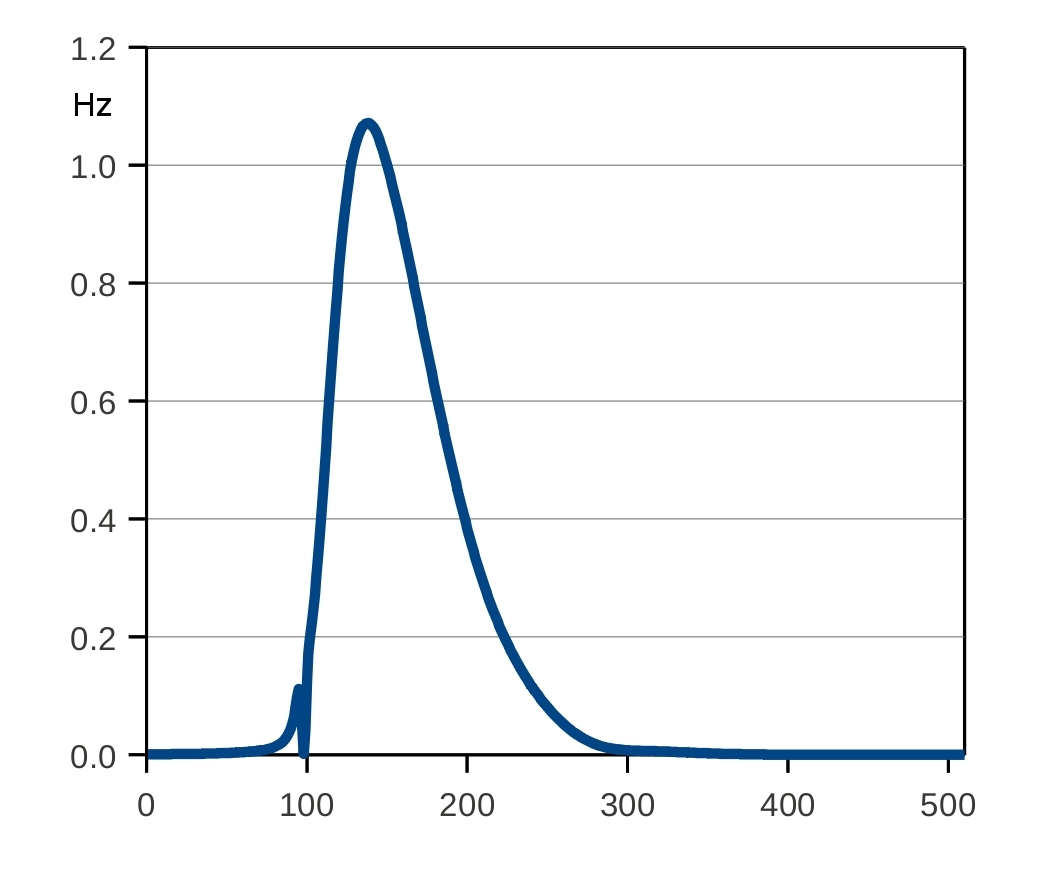}
\caption{\quad{}Inverse of $\Delta \tau_{\mathrm{rad}}$, computed from the initial static model.
Abscissa: gridpoints numbered from top to bottom, ordinate: $c_{\mathrm{cour}}/\Delta\tau_{\mathrm{rad }}$ in Hz. 
The H--ionisation zone is located at approximately 100, the He II--ionisation zone at approximately~300.}
\label{fig:iTau}
\end{figure}

For any study of the upper convection zone grid refinement is essential, as already pointed out in 
Sect.~\ref{subsec:resolving}. To exclude effects that might result from fringing due to the grid refinement zone in 
an experiment with one dimensional simulations the gridspacing was reduced over the whole domain. This improved
the accuracy of $Q_{\mathrm{rad}}$ and, consequently, the lightcurve smoothed out. This can be seen in Fig.~\ref{fig:1Dfine}
which also shows the results from a comparison run with the lower, original resolution. Both simulations were continued 
from the same initial model, an earlier state of the one-dimensional simulation on the coarser grid. Because of the 
higher spatial resolution the outgoing intensity becomes smaller: at the beginning the values are 30\% above the 
average which is reached after an initial relaxation phase. The jagged lightcurve obtained with the coarse grid is totally
unsuitable for comparisons with observations. This deficiency cannot be compensated by smoothing out the lightcurve
since even then the average would remain too large in comparison with the simulation with lower resolution
and the shape of the curve would still be different from a sufficiently resolved calculation (cf.\ Fig.~\ref{fig:1Dfine}).
The computational resources required for the sufficiently resolved case run are unacceptably high for applications in two 
dimensions. Hence, the grid refinement has been modified to make it partially adaptive. The region of interest 
is the region where we expect to find the 
H-ionisation zone. Below this zone the maximum possible timestep decreases swiftly (see Fig.~\ref{fig:iTau} for 
the inverse of $\Delta t_{\mathrm{rad}}$), therefore a finer grid can be justified only above this border. Since the 
location of the H-ionisation zone varies both in angular direction and with time a fixed lower boundary would invariably
lead to small timesteps, and thus we developed a co-moving lower boundary. The grid refinement is done over the 
entire angular direction, since otherwise artifacts occur.

\begin{figure*}
\centering
\includegraphics[width=\textwidth]{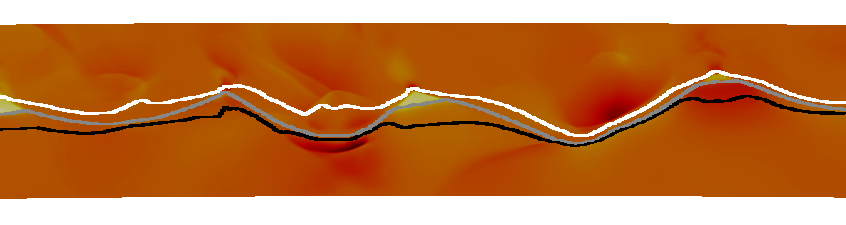}
\caption{\quad{}Increasing optical depth in the H ionisation zone on the fine grid (model 3). Approx. 1$^{\mathrm{o}}$ wide. White line: $\tau=1$; grey line: $\tau=10$; black line $\tau=100$; plotted over the convective flux $F_{\mathrm{c}} = \left( u_{r}-\overline{u}_{r}\right)\cdot\left(\rho h-\overline{\rho h}\right)$ where 
$h$ denotes the enthalpy.}
\label{fig:opacityH}
\end{figure*}

The most important region for grid refinement is located around the maximum of the temperature gradient. To 
prevent the code from finding an inappropriate maximum we also make sure that the local temperature is at least 
$T_{\mathrm{eff}}$ and the optical depth at least 100 at the lower border (black line in Fig.~\ref{fig:opacityH}).

\begin{figure*}
\centering
\includegraphics[width=\textwidth]{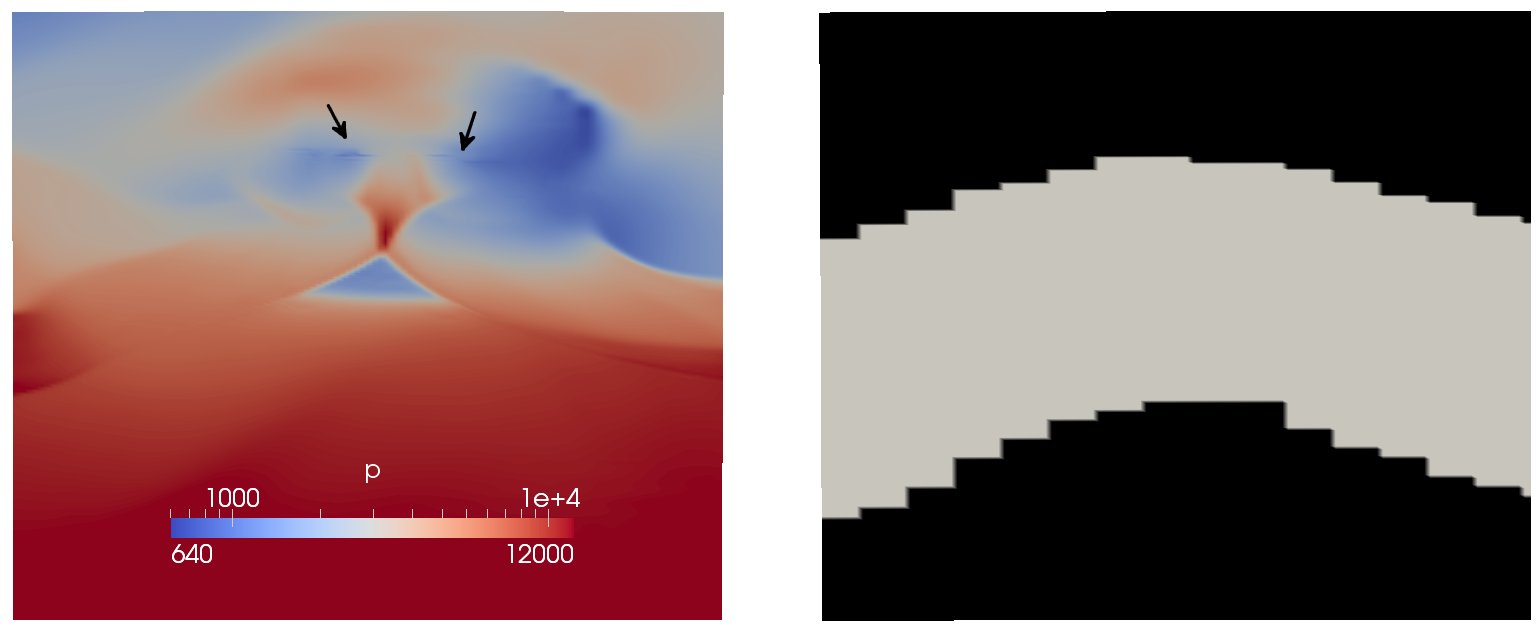}
\caption{\quad{}Grid refinement with upper limit on the ``inner region''. The left
panel shows pressure, the right panel indicates the location of the grid
refinement zone (grey: grid refinement zone, black: ghost cells).
Note the structures in pressure (given in dyn cm$^{-2}$) displayed in the left panel as
indicated by arrows which are just at the location of the upper
boundary indicated in the right panel.}    
\label{fig:upbound}
\end{figure*}

\begin{figure*}
\centering\includegraphics[width=\textwidth]{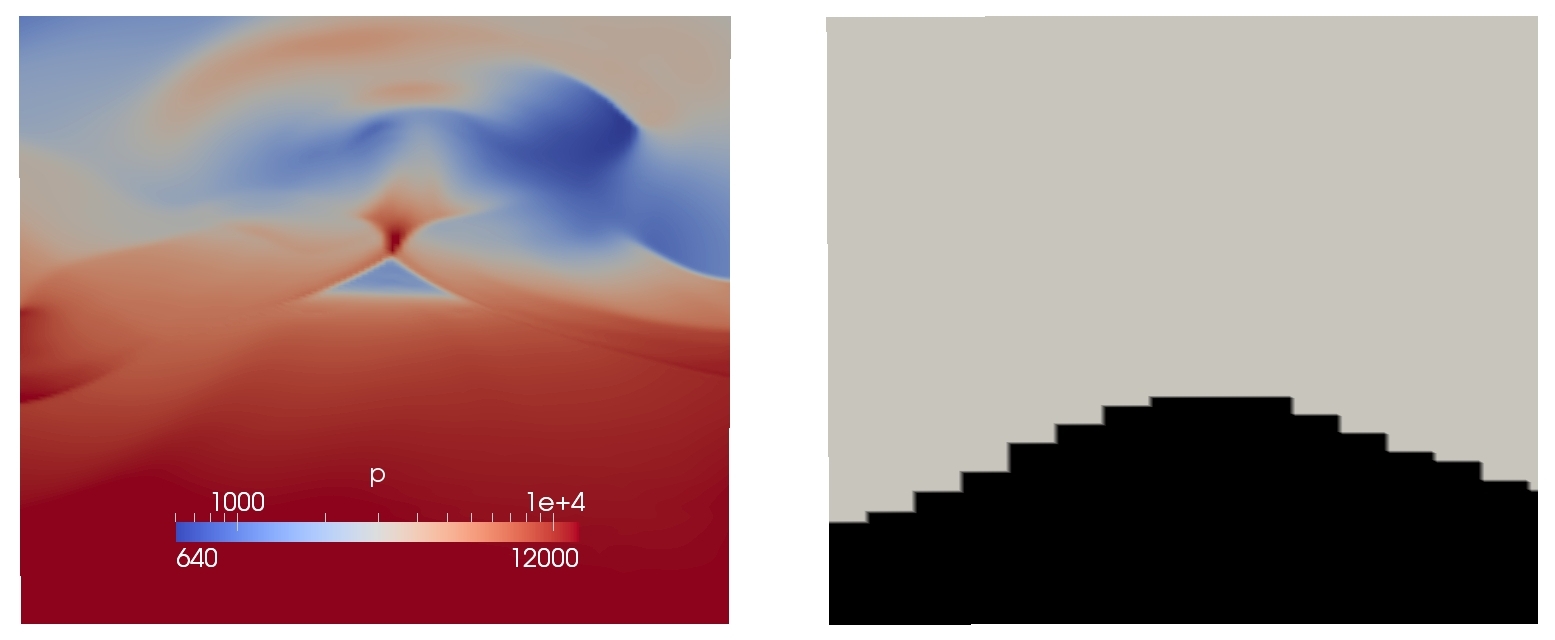}
\caption{\quad{}Grid refinement similar to Fig.~\ref{fig:upbound} but this time without upper limit on the refinement zone (``inner region''). 
Artifacts no longer occur.}    
\label{fig:noUpbound}
\end{figure*}

We now briefly describe the refinement procedure. We consider a region where the maximum
temperature gradient can be found and the grid refinement factors $\mathrm{gf}(r)$ and $\mathrm{gf}(\varphi)$
are specified. If the maximum temperature gradient is outside this region, the region has to be reset. In the first 
step the values of the physical quantities from the coarse grid are interpolated to the fine grid to obtain the starting 
state for the simulation. The step for the evolution in time is calculated for both grids. This yields the number $N$ 
of steps on the fine grid during one step on the coarse grid. Because the calculations are limited by $\Delta t_{\mathrm{rad}}$, 
we get $\mathrm{gf}(r)\leq N \leq \mathrm{gf}(r)^{2}$, where $\mathrm{gf}(r)$ is the radial grid refinement factor.
Since the domains are not identical, it is possible that $N$ is even smaller than $\mathrm{gf}(r)$, in which case
we set $N=\mathrm{gf}(r)$. Now the ``inner region'' is determined. It reaches from the top of the maximum region to 
a few points below the location of the maximum of the temperature gradient. Grid refinement is only done in this ``inner region''. For the
remaining points near the superadiabatic peak the values are interpolated from the coarse grid. The lower boundary 
of the ``inner region'' is determined in every step, so that the bottom-line moves with the lower boundary of the
actual H-ionisation zone. Artificial structures along this bottom line have not been observed by us in any of our simulations. 
Choosing, however, a top line in a similar manner we could not avoid getting artifacts: in the simulation shown in Fig.~\ref{fig:upbound} 
such a moving top boundary was still used while this was no longer the case for the simulation shown in 
Fig.~\ref{fig:noUpbound}. Note that in all figures all values outside the ``inner region'' are only interpolated from 
the coarse grid.

The time steps on the fine grid are done in exactly the same way as on the coarse grid. At the beginning of each 
step a time interpolation yields the values in the ghost cells at the top and bottom (black region in 
Fig.~\ref{fig:upbound} and Fig.~\ref{fig:noUpbound}) of the grid refinement zone. In angular direction they are 
obtained from periodic continuation. After the last step in the grid refinement region the data are projected back to the 
coarse grid in a conservative way.

For our two dimensional simulation (model 3) 3x4 additional grid points on each cell were used  in addition to subgrid 
scale modelling.  The resulting resolution in the critical region is $0.66~\mathrm{Mm}$ in radial and 
$1.16~\mathrm{Mm}$ in angular direction leading to an aspect ratio of $1:1.8$. But comparing to 
Fig.~\ref{fig:qrad} admittedly there are still only 2 to 3 points to resolve the temperature 
gradient.

\begin{figure*}
\centering
\includegraphics[width=\textwidth]{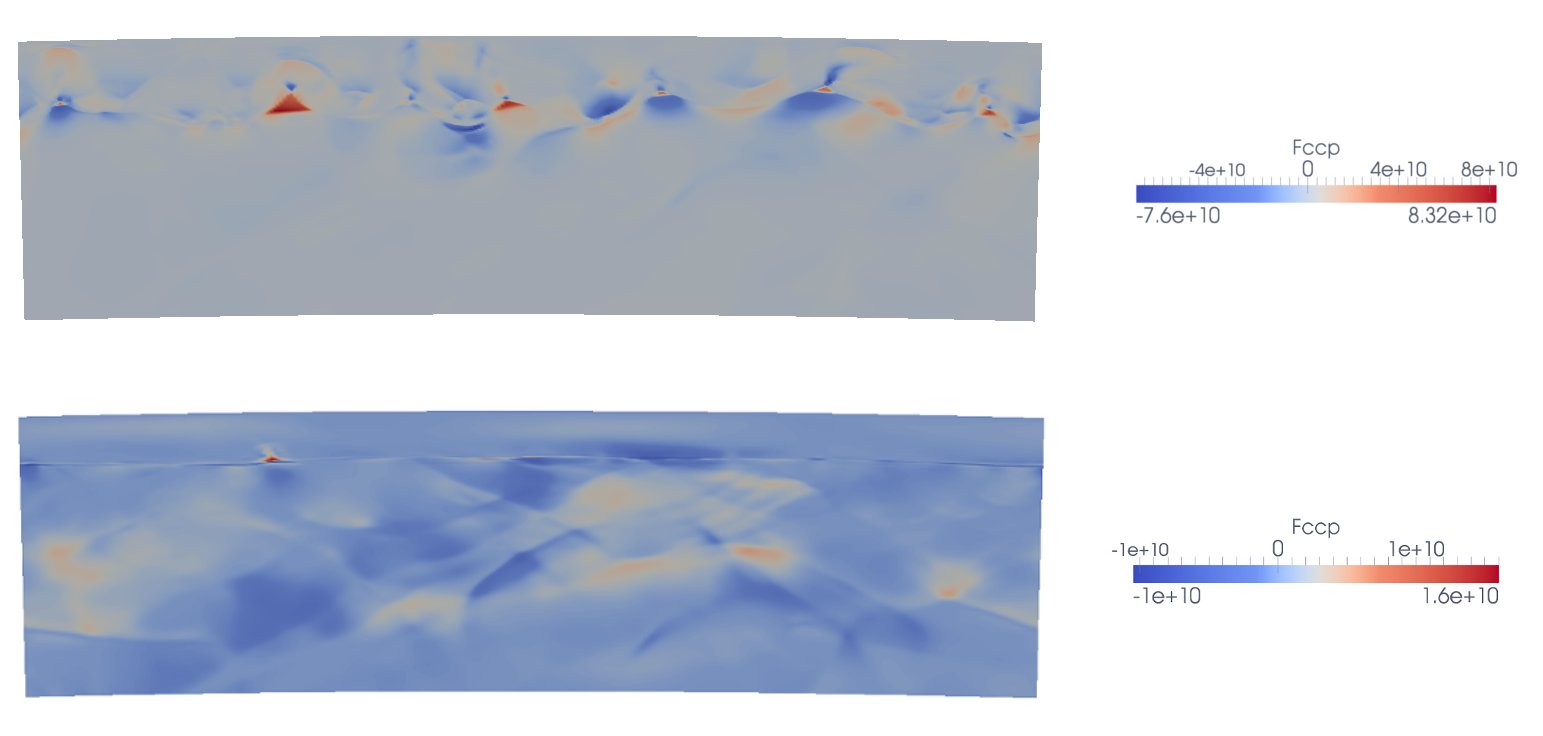}
\caption{\quad{}Convective flux (in erg s$^{-1}$ cm$^{-2}$) in the H-ionisation zone. The upper and lower panel 
depict the flux in the same area and at the same time according to the models with 
grid refinement (model~3, upper panel) and without (model~2, lower panel). The 
insufficiently resolved model~2 fails to properly develop the 
convection zone. Note that the same colours refer to drastically different values 
of the convective flux in both panels. Due to 
projection to the coarse grid for obtaining the upper panel, the refined model is not
displayed in full resolution.}  \label{fig:Gf-noGf} 
\end{figure*}

The uppermost panel in Fig.~\ref{fig:Gf-noGf}  is taken from the projection to the coarse grid and depicts the same area as the one 
below: the simulations both have an age of one day of stellar time. In both panels one 
can see the same maximum (red colour). However, not only is the convective flux already greater by a factor of 5, but there
are also more convection cells in model~3 in comparison with the lower resolution model~2 and the convection cells 
are of an unnatural shape in model~2. Also, the maximum Mach numbers are significantly lower in model~2. The properties of the 
atmosphere and the differences brought about by resolution will be discussed more closely in a subsequent paper. 
Already from these results, however, we can state that model~2 fails by far to properly represent the hydrogen ionisation zone
and the higher resolution model~3 is needed. 

\section{Discussion}  \label{sec:discussion}

\subsection{Comparisons and applications of the models}

Among the simulations listed in Table~\ref{tab:models} model~1 is particularly
useful for astrophysical studies. The simulation domain is sufficiently wide such 
that the He~{\sc ii} zone usually contains several up- and downflow structures. Its moderate
resolution is both its main strength and shortcoming. Because model~1 cannot resolve
the hydrogen ionisation zone, it is very robust. Strong shock fronts develop much more
rarely and, consequently, it has been easy to continue the simulation run over many
pulsation cycles to provide sound statistical averages. Naturally, with this model
studies on the convection-pulsation interaction have to be restricted to the 
role of the region around the zone of He~{\sc ii} ionisation. An in-depth study 
based on this model which investigates enthalpy fluxes and the pressure work 
done as a function of the pulsation phase will be presented in paper~II.

Model~2 mainly serves as a reference simulation for model~3, which is derived 
from it through grid refinement. It has the same resolution as model~1 and 
differs from it only by a more narrow opening angle, thus it has no advantage
over models~1 or model~3.

On the other hand, model~3 and model~4 provide two alternative ways of 
performing high resolution simulations which do resolve the H~{\sc i} and 
He~{\sc i} ionisation zone while at the same time they include the He~{\sc ii} 
ionisation zone and a sufficiently large region underneath it such that 
a meaningful and sufficiently accurate modelling of the radial pulsation 
is possible. As a consequence of its grid refinement model~3 has a wider range of
applications than model~4: it offers about the same resolution combined
with a much wider opening angle. Moreover, the maximum (resp.\ minimum)
aspect ratio of the grid cells throughout the simulation domain can be reduced
(kept closer to 1:1) while time step restrictions due to radiative diffusion in the 
layers below the superadiabatic layer can be eased as well. Finally, the He~{\sc ii} ionisation 
zone in model~4 is affected by periodicity effects along the azimuthal coordinate
which are caused by the necessarily small opening angle. This clearly demonstrates the 
advantages of grid refinement in combination with the spherical, radially stretched
and co-moving mesh used in all our simulations of Cepheids. From Fig.~\ref{fig:Gf-noGf}
it is evident that the dynamics at the optical surface of the Cepheid changes
considerably once sufficient resolution is achieved: sharp, pronounced shock fronts
appear which penetrate all the way up to the highest level of the photosphere.
These phenomena will be discussed in detail in paper~II. They appear in both
model~3 and model~4 and the similarity of these features demonstrates that 
the grid refinement does not introduce any artifacts (the large, cusp like features 
at the optical surfaces which can be found in Fig.~\ref{fig:upbound}--\ref{fig:Gf-noGf} 
are present in both model~3 and model~4 while they are hardly visible in model~2
in the lower panel of Fig.~\ref{fig:Gf-noGf}). Another consequence of 
successfully resolving the radiative cooling processes at the stellar surface is
a much more realistic, relatively smooth lightcurve as shown by Fig.~\ref{fig:1Dfine}
for the one-dimensional case. 

However, the new level of dynamics found in high resolution models also introduces 
new \textit{numerical} stability problems. Indeed, for a long period of time we have been 
quite content with the closed boundary conditions at the top of the domain as described 
in Sect.~\ref{initbound}. That changed once the much more vigorous convection pattern 
in the H~{\sc i} and He~{\sc i} ionisation zone was found in the simulations with grid 
refinement (see Fig.~\ref{fig:Gf-noGf}). When using the closed upper boundary condition 
as specified in Eq.~(\ref{equ:closed_top}) the faster flows and stronger shock waves 
generated in the highly resolved models no longer lead to a benign behaviour as observed 
for the weaker shocks that appear in low resolution models. As a result, high resolution 
models cannot successfully be continued over multiples of pulsation cycles. For such 
studies it is therefore necessary to develop appropriate variants of open boundary conditions.

\subsection{Overcoming the problem of small time steps}

To avoid unaffordable computational requirements and inefficient use of computational
resources high resolution simulations of the hydrogen ionisation zone, as discussed with
the examples of Fig.~\ref{fig:qrad} for the one-dimensional case, require {\emph{at least some}} 
terms to be treated with an implicit time integration method. The feasibility of such an approach 
for the case of a moving or even an adaptive grid has been shown in \cite{dorfi_99} for the 
$1\mbox{D}$ case. More recently, \cite{viall_11} have developed an implicit $2\mbox{D}$ code 
and provided test cases for problems in stellar physics. 

These codes treat both the Euler equation and the diffusive term in the energy equation implicitly. 
For Cepheid modelling in multidimensions the situation is different from what the latter authors 
seem to envisage. What matters here is the stringent limitation which is imposed by the diffusive 
term in the energy on the timestep, if explicit time integration is used, rather than, e.g, restrictions due 
to the high sound speed near the bottom of the simulation domain. These limitations, which are caused
by radiative diffusion, are due to unfavourable values of the physical parameters below the 
H+He~{\sc i} ionisation zone. On the other hand, the violent state of the atmosphere with strong shocks
ensuing once the resolution is sufficient precludes, in this case, useful {\it implicit} time marching for the 
Euler equations themselves. That would be advantageous only in the case of fast flows with slowly 
moving features or for a low Mach number flow. For the numerical simulation of Cepheids it should 
hence be sufficient to implicitly advance the diffusive terms in the energy equation in time in the 
relevant regions. A necessary task in this respect is the development of novel Runge-Kutta methods
which allow considerably larger time steps for such cases and replace the traditional schemes. Such
new methods have already been constructed and tested for their efficiency by \cite{hig_12}, where 
their applicability is demonstrated for the case of semiconvection, although the new schemes have 
actually been developed with Cepheids and A-type stars in mind. 

\section{Conclusions}  \label{sec:conclusions}

Setting the issue of strong shocks and obvious requirements such as the need 
for a moving grid and spherical geometry aside Cepheid modelling in 2D and 3D 
is made difficult due to extremely different spatial scales which must be resolved
simultaneously. This problem becomes evident only once realistic microphysics,
as characterised by the equation of state and radiative conductivities, combined
with realistic values for fluxes and a realistic stratification due to gravity 
are taken into account.

In our modified version of ANTARES we have tackled this challenge by introducing
a grid in a spherical coordinate system stretched along the radial direction. 
The resolution is maximised by letting the grid move with the mean velocity
of the uppermost domain layer and by introducing the possibility of grid refinement.
The latter is essential to resolve the superadiabatic layer and the dynamical processes
occurring in the photosphere such as sharp, fast moving shock fronts.

These changes in turn necessitated modifications to the WENO scheme used for
the spatial discretisation of the advection operator and the pressure gradients
in the dynamical equations. The new scheme discussed in this paper ensures 
conservation properties at the expense of a higher approximation order in smooth 
flow regions while it outperforms standard second order schemes due its smaller
error constant. For the same reasons the discretisation of diffusion operators 
had to be modified accordingly. These modifications allow to achieve a high 
spatial resolution at a fixed level of computational costs.

A mandatory feature of a simulation code used to perform realistic numerical 
simulations of Cepheids is a well scaling parallelisation. For this purpose
the radiative transfer scheme used in ANTARES was not only adapted to the 
new grid, but also implemented in a new way such that computational nodes 
can be distributed efficiently both along radial and azimuthal directions.
This minimises overheads due to data communication and allows a
cost-effective use of 256 processor cores even with grids consisting 
only of typically around 500 by 500 points.

Thanks to these features it has been possible to calculate a set 
of two-dimensional models which can be used to study the convection-pulsation
interaction of Cepheids for realistic stellar parameters and realistic microphysics.
A very high spatial resolution is needed to compute
reliable light-curves and reliably calculate the stratification in the superadiabatic
region. The grid refinement technique allows such resolution while covering
a wide simulation domain and at the same time avoid the introduction of
numerical artifacts.

For the time being two important problems remain on the technical and 
numerical side. Since unfavourably small timesteps are enforced by 
radiative diffusion below the H~{\sc i} and He~{\sc i} ionisation zone,
the simulations are currently very expensive. A suitable implicit time
integration procedure which eases these constraints is in development.
Secondly, for high resolution simulations, open upper boundary
conditions are required to allow stable integrations over many
pulsation cycles.

The models calculated thus far are already suited to investigate the
convection-pulsation interaction for the He~{\sc ii}-ionisation zone 
with realistic microphysics and to get first insights to the atmospheric 
structure of Cepheids with (grey) radiative transfer from $2\mbox{D}$ 
models. Such investigations will be presented in paper~II.

\section*{Acknowledgments}

This work has been supported by the Austrian Science Foundation, FWF grant P18224. FK acknowledges
support by the FWF grant P21742. We are thankful to G.~Houdek for supplying us with one-dimensional
starting models and to J. Ballot for carefully reading the manuscript and 
suggesting a number of improvements. Calculations have been performed at the VSC 
clusters of the Vienna universities.

\label{lastpage}


\begin{thebibliography}{99}
\bibitem[\protect\citeauthoryear{Aleshin}{1964}]{aleshin64} Aleshin, V.~I., 1964, Astronomicheskii Zhurnal, 41, 201
\bibitem[\protect\citeauthoryear{Baker \& Kippenhahn}{1962}]{bak62} Baker N., Kippenhahn R., 1962, ZA, 54, 114 
\bibitem[\protect\citeauthoryear{Bono \& Stellingwerf}{1994}]{bono_st94}Bono, G., \& Stellingwerf, R.~F., 1994, ApJS, 93, 233
\bibitem[\protect\citeauthoryear{Bono et al.}{1994}]{bono_2000}Bono, G., Caputo F., Cassini S., Marconi M.,Piersanti L., \&
Tornambe A., 2000, ApJ, 543, 971
\bibitem[\protect\citeauthoryear{Buchler}{1997}]{buch_rev97} Buchler, J.~R., 1997, in Variables Stars and the Astrophysical Returns of the Microlensing Surveys, eds. Ferlet, R., Maillard, 
J.-P., 181
\bibitem[\protect\citeauthoryear{Buchler 
\& Koll{\'a}th}{2000}]{buchko00} Buchler J.~R., Koll{\'a}th Z., 2000, NYASA, 898, 39
\bibitem[\protect\citeauthoryear{Buchler}{2009}]{buch_rev_09} Buchler, J.~R., 2009, American Institute of Physics Conference Series, 1170, 51
\bibitem[\protect\citeauthoryear{Buchler et al.}{1997}]{buch_koll97} Buchler, J.~R., Koll{\'a}th, Z., \& Marom, A., 1997, Ap\&SS, 253, 139
\bibitem[\protect\citeauthoryear{Carlson}{1963}]{carlson} Carlson, B.G., 1963, in Alder, B., Fernbach, S. (eds) Methods in Computational Physics, 1
\bibitem[\protect\citeauthoryear{Christy}{1962}]{christy62} Christy, R.~F., 1962, ApJ, 136, 887 
\bibitem[\protect\citeauthoryear{Christy}{1964}]{christy64} Christy, R.~F., 1964, Reviews of Modern Physics, 36, 555 
\bibitem[\protect\citeauthoryear{Cox}{1980}]{cox80} Cox A.~N., 1980, ARA\&A, 18, 15 
\bibitem[\protect\citeauthoryear{Cox}{1960}]{cox60} Cox J.~P., 1960, ApJ, 132, 594 
\bibitem[\protect\citeauthoryear{Cox et al.}{1966}]{cox66} Cox, J.~P., Cox, A.~N., Olsen, K.~H., King, D.~S., \& Eilers, D.~D., 1966, ApJ, 144, 1038 
\bibitem[\protect\citeauthoryear{Deupree}{1980}]{deup80} Deupree, R.~G., 1980, ApJ, 236, 225
\bibitem[\protect\citeauthoryear{Dorfi}{1999}]{dorfi_99} Dorfi E.~A., 1999, JCoAM, 109, 153 
\bibitem[\protect\citeauthoryear{Dorfi \& Feuchtinger}{1991}]{dorfi91} Dorfi, E.~A., \& Feuchtinger, M.~U., 1991, A\&A, 249, 417
\bibitem[\protect\citeauthoryear{Fedkiw et al.}{1998}]{fed1} Fedkiw, Merriman, Donat and Osher, July 1998, Progress in Numerical Solutions of Partial Difference Equations, Arachon, France (M.Hafez, ed.)
\bibitem[\protect\citeauthoryear{Ferguson et al.}{2005}]{aleFer} Ferguson, J.W., Alexander, D.R.. Allard, F., Barman, T., Bodnarik, J.G.,
Hauschildt, P.H., Heffner-Wong, A., Tamanai, A., 2005, ApJ 623, 585
\bibitem[\protect\citeauthoryear{Fokin}{1990}]{fok90} Fokin A., 1990, Ap\&SS, 164, 95
\bibitem[\protect\citeauthoryear{Freytag et al.}{2012}]{frey12} Freytag B., Steffen M., Ludwig H.-G., Wedemeyer-B{\"o}hm S., Schaffenberger W., Steiner O., 2012, JCoPh, 231, 919
\bibitem[\protect\citeauthoryear{Gastine \& Dintrans}{2011}]{gast_L11} Gastine, T., \& Dintrans, B., 2011, A\&A, 530, L7 
\bibitem[\protect\citeauthoryear{Geroux \& Deupree}{2011}]{ger11} Geroux, C.~M., \& Deupree, R.~G., 2011, ApJ, 731, 18
\bibitem[\protect\citeauthoryear{Grevesse \& Noels}{1993}]{GN_1993} Grevesse, N., \& Noels, A., N. Prantzos, E. Vangioni-Flam and 
M. Cass\'e (eds.), 1993, 
     Symposium in Honour of Hubert Reeves' 60th birthday: Origin and evolution of the elements, pp.15
\bibitem[\protect\citeauthoryear{de Grijs}{2011}]{degrijs11} de Grijs, R., 2011, 
     An Introduction to Distance Measurement in Astronomy (Wiley)
\bibitem[\protect\citeauthoryear{Happenhofer et al.}{2011}]{happ11} Happenhofer, N., Grimm-Strele, H., Kupka, F., L\"ow-Baselli, B.,
   \& Muthsam, H., 2011, to appear in J.\ Comput.\ Phys.\ (preprint available at http://arxiv.org/abs/1112.3507)    
\bibitem[\protect\citeauthoryear{Harten et al.}{1987}]{harten}Harten, A., Enquist, B., Osher, S. \& Chakravarthy, S.~R., 1987, 
 J.\ Comput.\ Phys.\ 71, 23
\bibitem[\protect\citeauthoryear{Hertzsprung}{1926}]{hertz26} Hertzsprung, E.\ 1926, Bull. Astron. Inst. Netherlands, 3, 115
\bibitem[\protect\citeauthoryear{Higueras et al.}{2012}]{hig_12} Higueras, I., Happenhofer, N., Koch, O., Kupka, F. \ 2012,
submitted to SIAM J. Scientific Computing; preprint at www.asc.tuwien.ac.at/preprint/2012/asc14x2012.pdf
\bibitem[\protect\citeauthoryear{Iglesias \& Rogers}{1996}]{opalOpac}Iglesias, C.A. \& Rogers, F.J., 1996. ApJ 464, 943
\bibitem[\protect\citeauthoryear{Jiang \& Shu}{1996}]{jiang}Jiang, G.~S. \& Shu, C.-W., 1996, J.\ Comput.\ Phys., 126, 202
\bibitem[\protect\citeauthoryear{Keller}{2008}]{keller08} Keller 
S.~C., 2008, ApJ, 677, 483
\bibitem[\protect\citeauthoryear{Koch et al.}{2010}]{koch10} Koch, O., Kupka, F., L\"ow-Baselli, B., Mayrhofer, A., \& Zaussinger, F.,
ASC Report 32/2010, ISBN 978-3-902627-03-2, Institute for Analysis and Scientific Computing, Vienna University of Technology, Wien 
(http://www.asc.tuwien.ac.at/preprint/2010/asc32x2010.pdf)
\bibitem[\protect\citeauthoryear{Kunasz \&  Auer}{1988}]{kun_auer} Kunasz, P. and Auer, L.~H., 1988, JQSRT, 39, 67
\bibitem[\protect\citeauthoryear{Kupka, Ballot, \& Muthsam}{2009}]{kup09} Kupka F., Ballot J., Muthsam H.~J., 2009, CoAst, 160, 30
\bibitem[\protect\citeauthoryear{Lilly}{1962}]{lilly}Lilly, D.~K., 1962, Tellus, 14 (2), 148
\bibitem[\protect\citeauthoryear{Liu et al.}{1994}]{liu1}Liu, X., Osher, S., Chan, T., 1994, J.\ Comput.\ Phys., 115, 200
\bibitem[\protect\citeauthoryear{Marconi}{2009}]{marc_rev09}  Marconi, M., 2009, American Institute of Physics Conference Series, 1170, 223
\bibitem[\protect\citeauthoryear{Merriman}{2003}]{merr_03}  Merriman, B., 2003, J.\ Scientific Computing, 19, 309 
\bibitem[\protect\citeauthoryear{Muthsam et al.}{2010}]{muth_na10} Muthsam, H.~J., Kupka, F., L{\"o}w-Baselli, B., Obertscheider, C., Langer, M. \& Lenz, P., 2010, NewA, 15, 460
\bibitem[\protect\citeauthoryear{Muthsam et al.}{2012}]{muth_mnras12} Muthsam, H.~J., Mundprecht, E., Kupka, F., , 2012, in preparation
\bibitem[\protect\citeauthoryear{Nordlund}{1982}]{nord82} Nordlund A., 1982, A\&A, 107, 1
\bibitem[\protect\citeauthoryear{Olson \&  Kunasz}{1987}]{kun_olson} Olson, L.~H and  Kunasz, P., 1987, JQSRT, 38, 325
\bibitem[\protect\citeauthoryear{Rogers et al.}{1996}]{opalEos} Rogers, F.~J., Swenson, F.~J., and Iglesias, C.~A., 1996, ApJ 456, 902
\bibitem[\protect\citeauthoryear{Rogers 
\& Nayfonov}{2002}]{rogers_nay_02} Rogers F.~J., Nayfonov A., 2002, ApJ, 576, 1064 
\bibitem[\protect\citeauthoryear{Rosenthal et al.}{1999}]{rosen99} Rosenthal, C.~S., Christensen-Dalsgaard, J., Nordlund, {\AA}., 
Stein, R.~F., Trampedach, R., 1999, A\&A, 351, 689
\bibitem[\protect\citeauthoryear{Shu \& Osher}{1988}]{shu1}Shu,C.-W., Osher, S., 1988, J.\ Comput.\ Phys.\ 77, 439
\bibitem[\protect\citeauthoryear{Shu}{1997}]{shu2}Shu,C.-W., 1997,  Technical report NASA CR-97-206253 ICASE Report No. 97-65, Institute for Computer Applications in Science and Engineering
\bibitem[\protect\citeauthoryear{Shu}{2003}]{shuGal}Shu,C.-W., 2003, International Journal of Computational Fluid Dynamics, 17:2, 107
\bibitem[\protect\citeauthoryear{Smagorinsky}{1963}]{smagorinsky}Smagorinsky, J., 1963,  Mon. Weather Rev., 91, 99 
\bibitem[\protect\citeauthoryear{Smolec}{2008}]{smol08}  Smolec, R., 2008, CoAst, 157, 149
\bibitem[\protect\citeauthoryear{Smolec \& Moskalik}{2008}]{smolec_mo08} Smolec, R., \& Moskalik, P., 2008, Acta Astron., 58, 193
\bibitem[\protect\citeauthoryear{Spiegel}{1957}]{spiegel} Spiegel, E.~A., 1957, ApJ 126
\bibitem[\protect\citeauthoryear{Tuggle \& Iben}{1973}]{tuggle_iben73}  Tuggle, R.~S., \& Iben, I., Jr., 1973, ApJ, 186, 593
\bibitem[\protect\citeauthoryear{Viallet, Baraffe, 
\& Walder}{2011}]{viall_11} Viallet M., Baraffe I., Walder R., 2011, A\&A, 531, A86
\bibitem[\protect\citeauthoryear{Zhevakin}{1953}]{zhev53}  Zhevakin, S.~A., 1953, Russ. A.J., 30, 161
\end{thebibliography}
\end{document}